\documentclass{JHEP3}

\usepackage{epsfig}
\usepackage{amsfonts}
\usepackage{amssymb,amsmath}

\newcommand{\mbf}[1]{{\boldsymbol {#1} }}
\newcommand{\complex}{{\mathbb C}} 
\newcommand{\zed}{{\mathbb Z}} 
\newcommand{\nat}{{\mathbb N}} 
\newcommand{\real}{{\mathbb R}} 
\newcommand{\rat}{{\mathbb Q}} 
\newcommand{\torus}{{\mathbb T}}
\def\e{{\,\rm e}\,}

\newcommand{\id}{{1\!\!1}}

\def\ii{{\,{\rm i}\,}}
\def\dd{{\rm d}}
\def\DD{{\rm D}}

\def\beq{\begin{equation}}
\def\eeq{\end{equation}}
\def\bea{\begin{eqnarray}}
\def\eea{\end{eqnarray}}
\def\bd{\begin{displaymath}}
\def\ed{\end{displaymath}}

\newcommand{\be}{\begin{equation}}
\newcommand{\ee}{\end{equation}}

\newcommand\fverb{\setbox\pippobox=\hbox\bgroup\verb}
\newcommand\fverbdo{\egroup\medskip\noindent%
                        \fbox{\unhbox\pippobox}\ }
\newcommand\fverbit{\egroup\item[\fbox{\unhbox\pippobox}]}
\newbox\pippobox

\preprint{{\tt HWM-05-15} \ \  {\tt EMPG-05-11}\\
\hepth{0506016}\\ June 2005}

\title{Morita Duality and Noncommutative Wilson Loops\\ in Two Dimensions}

\author{Michele Cirafici$^a$,
Luca Griguolo$^b$, Domenico Seminara$^{c}$ and Richard J. Szabo$^{a}$ \\
$^a$ Department of Mathematics, Heriot-Watt University\\
Colin Maclaurin Building, Riccarton, Edinburgh EH14 4AS, UK\\
$^b$ Dipartimento di  Fisica, Universit\`a  di Parma,
INFN Gruppo Collegato di Parma\\
Parco Area delle Scienze 7/A, 43100 Parma, Italy\\
$^c$ Dipartimento di Fisica, Polo Scientifico Universit\`a di
Firenze, INFN Sezione di Firenze
Via  G. Sansone 1, 50019 Sesto Fiorentino, Italy\\
\email{M.Cirafici@ma.hw.ac.uk , griguolo@fis.unipr.it ,
seminara@fi.infn.it , R.J.Szabo@ma.hw.ac.uk}}

\abstract{We describe a combinatorial approach to the analysis of the
  shape and orientation dependence of Wilson loop observables on
  two-dimensional noncommutative tori. Morita equivalence is used to
  map the computation of loop correlators onto the combinatorics of
  non-planar graphs. Several strong nonperturbative evidences of symmetry
  breaking under area-preserving diffeomorphisms are thereby
  presented. Analytic expressions for correlators of Wilson loops with
infinite winding number are also derived and shown to agree with
results from ordinary Yang--Mills theory.}

\begin{document}

\section{Introduction}

The calculation of Wilson loop obervables is
an ongoing area of activity in the study of noncommutative gauge
theories (see~\cite{KSc1}--\cite{Sz1} for reviews). Early calculations
performed in the corresponding dual supergravity
theories~\cite{Alishahiha:1999ci} found that, in even spacetime
dimension and for maximal rank noncommutativity, large area Wilson
loop correlators behave exactly as their commutative counterparts up
to a rescaling of the Yang--Mills coupling constant, while
noncommutative effects dominate small area loops. In contrast,
numerical studies based on twisted reduced models in two
dimensions~\cite{BHN1} have revealed that small noncommutative Wilson
loops follow an area law behaviour while large loops
become complex-valued with a phase that rises linearly with the
area. Numerical results in four dimensions and for non-maximal
rank noncommutativity are qualitatively similar~\cite{BBHNSV1}.

In this paper we will analyse some nonperturbative properties of
Wilson loops in two-dimensional noncommutative gauge theory. The
partition function and open Wilson line observables in this theory
have been computed nonperturbatively in terms of instanton
expansions~\cite{GSV1}--\cite{GrigSem1} which are
manifestly invariant under gauge Morita equivalence and area-preserving
diffeomorphisms of the spacetime. In marked contrast, closed Wilson
line observables have thus far only been amenable to a variety of
perturbative studies~\cite{BNT1}--\cite{BDPTV1}. Foremost among the
interesting effects that have been unveiled is the loss of invariance
under area-preserving diffeomorphisms of the two-dimensional
spacetime~\cite{ADM1,BDPTV1}. Commutative Wilson loop correlators in
two dimensions are well-known~\cite{Witten:1991we} to be independent
of the shape of the contour on which they are defined and
depend only on the area enclosed by the loop. However, in
noncommutative gauge theory on $\real^2$ the loop correlators depend
on the path shape~\cite{ADM1,BDPTV1}. Thus, for example, one obtains
different correlation functions associated to a circular loop and a
square loop which encircle the same area. The simplest way to
understand the violation of this invariance is through the
noncommutative loop equation~\cite{ADM1}, which relates an
infinitesimal variation in the loop geometry of a closed Wilson line
correlator to a non-vanishing correlation function of open Wilson
lines. This symmetry breaking may be related to the
fact that, unlike its commutative version, the lattice regularization
of the noncommutative gauge theory~\cite{AMNS1,Ambjorn:2000cs} is not
invariant under subdivision of plaquettes which have long-ranged
interactions with one another. The standard Gross--Witten
reduction~\cite{GW1} breaks down due to UV/IR mixing in this
case~\cite{PanSz3}. On the other hand, it is expected~\cite{BDPTV1}
that at least an $SL(2,\real)$ subgroup of area-preserving
diffeomorphisms remains a symmetry of the quantum averages in
perturbation theory.

In the following we will study the shape dependence of loop
correlators from a nonperturbative perspective. Our fundamental point
of view will be to look at Morita equivalent formulations of the gauge
theory on a two-dimensional noncommutative torus. Since rational
noncommutative Yang--Mills theory is equivalent to ordinary
Yang--Mills theory on a torus, one would naively expect that in this
case Wilson loop correlators are shape-independent. By continuity one
could then try to extrapolate this result to the irrational
noncommutative torus and by decompactification even to the
noncommutative plane. The reason this argument breaks down is that
closed Wilson lines, unlike the open ones, have a very intricate
transformation property under Morita equivalence. The Morita dual of a
closed simple curve can be a very complicated loop with many
self-intersections and windings around itself. We describe these
transformations in detail, and show that Morita equivalence maps a
simple noncommutative Wilson loop on the torus into a non-planar graph
realizing a triangulation of the dual torus. The problem of computing
the loop correlator is in this way mapped onto a combinatorial
problem. Loops which enclose the same area but have a different shape
can yield topologically inequivalent graphs and hence different
correlation functions. This is in fact also true of loops which
differ only in their relative orientation, a feature which
distinguishes observables on the torus from those on the plane which
are rotationally invariant~\cite{BDPTV1}.
The loss of invariance under area-preserving diffeomorphisms from this
perspective is then attributed to the different graph combinatorics
induced by contours of varying shape. The spacetime transformations
which leave a given loop correlator invariant are determined by the
automorphism group of the non-planar graph induced by the contour
under Morita equivalence.

The organisation of this paper is as follows. In Section~\ref{MEWL} we
review some aspects of Morita equivalence and spell out in detail how
it acts on closed Wilson lines. In Section~\ref{DLCR} we present
various explicit constructions and calculations in rational
noncommutative gauge theory on the torus. In this case the Morita
dual gauge theory can be taken to be ordinary Yang--Mills theory, in
which we can perform calculations of self-intersecting loop
correlators using combinatorial techniques. Our explicit
nonperturbative expressions indeed do suggest the claimed shape
dependence, and we present various supporting arguments for this
claim. In Section~\ref{DLCI} we make some remarks concerning
irrational noncommutative gauge theories. Although we cannot make
progress with analytical determinations of loop correlators in this
case, we can give a heuristic picture of irrational noncommutative
Wilson loops as infinitely wound and self-intersecting contours in
some dual gauge theory. In Section~\ref{Wilsoninf} we then give an
explicit realization of this infinite winding property, and derive a
nonperturbative expression for the Wilson loop correlator on the
noncommutative plane in this case which coincides with the result of
resumming commutative planar diagrams in perturbation theory. Finally,
in Section~\ref{Summary} we summarize our findings and make some
further remarks about the relation between our nonperturbative
approach and existing perturbative calculations.

\section{Morita Equivalence of Wilson Loops\label{MEWL}}

In this section we will recall some basic features of two
dimensional noncommutative Yang--Mills theory. When this theory is
defined on a noncommutative torus, there exists a powerful tool to
perform explicit computations called Morita equivalence. This is
a duality that relates observables in the noncommutative gauge theory to
observables in a dual gauge theory. When the noncommutativity parameter
is a rational number, the equivalence can be arranged so that the dual
theory is a commutative gauge theory and
the standard techniques of ordinary Yang--Mills theory in two
dimensions can be applied to compute quantum correlation functions.

Consider $U(1)$ Yang--Mills theory
 defined on a square noncommutative torus $\mathbb{T}^2_{\theta}$ with
 noncommutativity parameter $\theta\in\real$, so that $[ x^1 , x^2 ]_\star=
 \ii \theta$ with $\mbf x=(x^1,x^2)$ local coordinates on the
 torus. The radius of $\mathbb{T}^2_{\theta}$ is $r'$ so that one has
 the identifications
\begin{equation}
x^\mu\sim x^\mu+ 2 \pi\,r' \ , \quad \mu=1,2 \ .
\end{equation}
While the main features below will hold for general $\theta$, we will
mostly refer to the gauge theory with rational-valued dimensionless
noncommutativity parameter of the form $\Theta = \frac{\theta}{2 \pi\,
r'{\,}^2} = - \frac{c}{N}$ with $c,N$ relatively prime positive
integers. The Yang--Mills action is given by
\begin{equation} \label{NCaction}
S_{\rm NCYM}[\mathcal{A}]
= \frac{1}{2g'{\,}^2}\,\int_{\torus_\theta^2}\dd^2\mbf x~ \left(
\mathcal{F} + \Phi \right)^2 \ ,
\end{equation}
where the Yang--Mills field strength
\begin{equation}
\mathcal{F}= \partial_1\mathcal{A}_2 -
\partial_2\mathcal{A}_1- \ii \left( \mathcal{A}_1 \star \mathcal{A}_2
- \mathcal{A}_2 \star \mathcal{A}_1\right)
\end{equation}
with $\partial_\mu=\partial/\partial x^\mu$ is defined in terms of
the abelian noncommutative gauge field $\mathcal{A}_{\mu}$ which has a
Fourier series expansion
\begin{equation}
\mathcal{A}_{\mu}(\mbf x) = \sum_{{\mbf q}\in\zed^2}\,a_{{\mbf q};\mu}~
\e^{-\ii{\mbf q} \cdot {\mbf x}/r'} \ , \quad a_{{\mbf
    q};\mu}\in\complex \ .
\end{equation}
Hermiticity of the gauge field requires $a_{-\mbf
  q;\mu}=\overline{a_{\mbf q;\mu}}$. The star-product of fields is
defined as
\begin{equation}
(f\star g)(\mbf x) = f(\mbf x)~\exp \left(\mbox{$\frac{\ii\theta}{2}~
\epsilon^{\mu\nu}~\overleftarrow{\partial_\mu}~\overrightarrow{\partial_\nu}
$}\,\right)~g(\mbf x) \ ,
\end{equation}
and we have introduced a constant abelian background flux $\Phi$.

Observables of noncommutative gauge theories are given by closed and
open Wilson lines~\cite{IIKK1}. In this paper we will focus
only on closed paths, whose corresponding Wilson lines are defined as
\begin{equation}
\label{NCwloop} \mathcal{O}_{\star} \left( \mathcal{C} \right) =
\int_{\torus_\theta^2} \mathrm{d}^2\mbf x ~\mathcal{U} \left( x ;
\mathcal{C} \right)
\end{equation}
where $\mathcal{C}$ is a closed contour on $\mathbb{T}^2_{\theta}$
with embedding
$\mbf\xi=(\xi^1,\xi^2):[0,1]\to\torus_\theta^2$,
$\xi^\mu(0)=\xi^\mu(1)$ and
\bea \label{NCPexp}
\mathcal{U} \left( x ; \mathcal{C} \right) &=&
\mathrm{P}_{\star}~\exp \left(\ii \oint_{\mathcal{C}}
\mathrm{d}\xi^{\mu}~\mathcal{A}_{\mu} \left(\mbf x + \mbf\xi \right)
\right) \nonumber\\ &=&1 + \sum_{n=1}^{\infty} \ii^n\,\int_{0}^{1}
\mathrm{d}s_1~\int_{0}^{s_1} \mathrm{d}s_2~\cdots
\int_{0}^{s_{n-1}} \mathrm{d}s_{n}~\dot\xi^{\mu_1}(s_1)\,
\dot\xi^{\mu_2}(s_2) \cdots \dot\xi^{\mu_n}(s_n)\nonumber\\ && \quad
\times\, \mathcal{A}_{\mu_1} \bigl(\mbf x + \mbf\xi(s_1)\bigr) \star
\mathcal{A}_{\mu_2} \bigl(\mbf x + \mbf\xi(s_2)\bigr) \star \cdots \star
\mathcal{A}_{\mu_n} \bigl(\mbf x + \mbf\xi(s_n)\bigr)
\eea
with $\dot\xi^\mu(s)=\dd\xi^\mu(s)/\dd s$ is the noncommutative
holonomy. The technique we will employ to compute noncommutative
Wilson loop correlators is to implement Morita equivalence at the
level of these observables in the case of a rational noncommutativity
parameter where the target dual gauge theory is commutative, and use
the known techniques \cite{Witten:1991we} to compute the
correlators in ordinary Yang--Mills theory. As we will see in the
following, this procedure, though naively well-defined, is full of
subtleties that need to be dealt with.

Generally, gauge Morita equivalence is a map between the
noncommutative gauge theory with action given by (\ref{NCaction}) and a
$U(N)$ noncommutative gauge theory on another torus
$\mathbb{T}^2_{\tilde\theta}$ with $m$ units of background magnetic
flux whose parameters are related to those of the original
theory by the action of an $SL(2,\mathbb{Z})$ duality
group~\cite{Schwarz1}. Explicitly, the parameters of the two gauge
theories are related as
\begin{eqnarray}
\left( \begin{array}{c} m \\ N \end{array} \right) & = &
\left( \begin{array}{cc} a~ &~ b \\ c~ & ~d \end{array} \right)
\left( \begin{array}{c} 0 \\ 1 \end{array} \right) \ , \nonumber\\
\tilde\Theta & = &\frac {c + d\,\Theta}{a + b\,\Theta} \ , \nonumber\\
\tilde r& = & |a + b\,\Theta|\,r \ , \nonumber \\
\tilde g^2 & = & (a + b\,\Theta)\,g^2 \ , \nonumber\\
\tilde\Phi&=&(a+b\,\Theta)^2\,\Phi-\frac{b\,(a+b\,\Theta)}
{2\pi\,r^2} \ ,
\label{Moritagen}\end{eqnarray}
where $a,b,c,d\in\zed$ satisfy the Diophantine relation
$a\,d-b\,c=1$. These relations guarantee invariance of the Yang--Mills
action under the Morita transformation.

For the particular case in
which the original $U(1)$ gauge theory has a rational-valued
noncommutativity parameter $\Theta= - \frac{c}{N}$, the
$SL(2,\mathbb{Z})$ element above (having $b=m,d=N$) yields a vanishing
$\tilde\Theta$ and the dual theory is a commutative $U(N)$ Yang--Mills
theory with coupling constant
\begin{equation} g^2 = \frac{g'{\,}^2}{N} \end{equation}
defined on a torus of radius
\begin{equation} \label{transR2}
r= \frac{r'}{N}
\end{equation}
with a non-trivial magnetic flux. This relation will play an important
role in the following, as it implies that the target torus is smaller than
 the original one since its area shrinks by a factor $N^2$. When
 considering Wilson loops on the torus, we will have to deal with
 this shrinking. We will be primarily concerned with
 this form of the duality, since some explicit nonperturbative
 computations can be done on the commutative torus. Moreover,
 since any irrational number is the limit of an infinite
 sequence of rational numbers, we expect that the results obtained
 in this way hold at general values of $\Theta$, or equivalently
 that they are continuous functions of the noncommutativity parameter. In the
 following we will conventionally refer to the $U(1)$
 noncommutative gauge theory with primed variables and to its
 commutative Morita dual gauge theory with unprimed variables.

We will make use of Morita duality to compute
correlators in the noncommutative gauge theory by computing their
commutative counterparts. This idea has been exploited
in~\cite{GSV1}--\cite{GrigSem1} and it enables one to perform
calculations very explicitly. To complete this program, we have to
exhibit the transformation law of the observables under Morita
duality. This problem has been solved
in~\cite{Ambjorn:2000cs,Guralnik:2001pv}. For example, to the operator
(\ref{NCwloop}) we associate the commutative Wilson loop
\begin{equation} \label{Cwloop}
\mathcal{O} \left( \mathcal{C} \right) =
\int_{\torus^2}\mathrm{d}^2\mbf x~
\mathrm{Tr}^{~}_N~\mathrm{P}~\exp\left( \ii \oint_{\mathcal{C}}
\mathrm{d}\xi^{\mu}~A_{\mu} \left( \mbf x + \mbf\xi \right) \right)
\end{equation}
where $\mathrm{Tr}^{~}_N$ is the trace in the fundamental
representation of the $U(N)$ gauge group, and the commutative
path-ordering operator P is defined as the
analog of (\ref{NCPexp}) with star-products replaced by
ordinary matrix products and the noncommutative gauge fields
$\mathcal{A}_{\mu}$ by their commutative counterparts $A_{\mu}$.
The identification of the observables is then given by
\begin{equation} \label{equivloops}
\mathcal{O}_{\star} \left( \mathcal{C} \right) = N\,\mathcal{O}
\left( \mathcal{C} \right) \ .
\end{equation}
As this equation is of fundamental importance to us, let us briefly
review its derivation following~\cite{Guralnik:2001pv}.

Consider commutative pure $U(N)$ gauge theory on $\mathbb{T}^2$ with
$m$ units of background magnetic flux. A non-trivial flux
implies that the gauge fields obey twisted boundary conditions on the
torus. They are solved by the Fourier expansions
\begin{equation} \label{CA}
A_{\mu} ({\mbf x}) = \sum_{{\mbf q}\in\zed^2}\,a_{{\mbf q};\mu}~
Q^{-c\,q^1}\,P^{q^2}~\e^{-\pi\ii c\,q^1\,q^2/N}~
\e^{-\ii {\mbf q} \cdot {\mbf x}/N\,r}
\end{equation}
where $P$ and $Q$ are the usual shift and clock matrices of rank $N$
which obey the commutation relation $P\,Q =\e^{2 \pi \ii/N}\,Q\,P$.
The $n^{\rm th}$ term in the expansion of the Wilson loop observable
$N\,\mathcal{O}(\mathcal{C})$ given by (\ref{Cwloop}) then takes the
form
\begin{eqnarray} \label{step1}
&& \ii^n\,N\,\int_{0}^{2 \pi\,r} \mathrm{d}x^1~\int_0^{2\pi\,r}
\mathrm{d}x^2~\prod_{i=1}^n\,
\int_{0}^{s_{i-1}} \mathrm{d}s_i~\sum_{{\mbf q}_i\in\zed^2}
\dot\xi^{\mu_i}(s_i)\,a_{{\mbf q}_i; \mu_i} \nonumber\\
&& \qquad\times~\mathrm{Tr}^{~}_N
\left( Q^{-c\,q_1^1}\,P^{q_1^2} \cdots Q^{-c\,q_n^1}\,P^{q_n^2}
\right)~\prod_{i=1}^n\e^{-\pi \ii c\,q_i^1\,q_i^2/N}~\e^{-\ii {\mbf q}_i
\cdot ({\mbf x} + \mbf\xi(s_i) )/N\,r}
\end{eqnarray}
where we have defined $s_0=1$. By using the commutation properties of
the clock and shift matrices it follows that
\begin{equation}
\mathrm{Tr}^{~}_N\left( Q^{-c\,q_1^1}\,P^{q_1^2} \cdots Q^{-c\,q_n^1}\,
P^{q_n^2} \right) = \mathrm{Tr}^{~}_N \left( Q^{-c\,(q_1^1 + \cdots +
q_n^1) }\,P^{q_1^2 + \cdots + q_n^2} \right)~\e^{- \frac{2 \pi \ii c}N\,
\sum\limits_{i>j} q_i^1\,q_j^2} \ .
\end{equation}
The trace on the right-hand side of this equation vanishes unless
$q_1^\mu + \cdots + q_n^\mu = q^\mu\,N$ for some integers $q^\mu$. If
these conditions are satisfied, then since $P^N=Q^N=\id_N$ the trace
is equal to $N$. Finally, due to these momentum constraints the
integrals over $\torus^2$ give Kronecker delta-functions, and we can
thereby formally rewrite (\ref{step1}) as
\begin{eqnarray} \label{step2}
&& \ii^n\,\int_{0}^{2 \pi\,r'} \mathrm{d}x^1~\int_0^{2\pi\,r'}
\mathrm{d}x^2~\prod_{i=1}^n\,
\int_{0}^{s_{i-1}} \mathrm{d}s_i~\sum_{{\mbf q}_i\in\zed^2}
\dot\xi^{\mu_i}(s_i)\,a_{{\mbf q}_i; \mu_i}~\e^{-\ii{\mbf q}_i\cdot
\mbf\xi(s_i)/r'}~
\e^{-\pi \ii c\,q_i^1\,q_i^2/N}  \nonumber\\ &&  \qquad\times~\e^{- 
\frac{2 \pi \ii c}N\,
\sum\limits_{i>j} q_i^1\,q_j^2}~\e^{-\ii(q_1^1 + \cdots
+ q_n^1) x^1/r'}\star \e^{-\ii(q_1^2 +\cdots + q_n^2) x^2/r'} \ .
\end{eqnarray}
By repeatedly using the properties of the star-product, it is
straightforward to recast (\ref{step2}) into the form
\begin{eqnarray}
&& \ii^n\,\int_{0}^{2 \pi\,r'} \mathrm{d}x^1~\int_0^{2\pi\,r'}
\mathrm{d}x^2~\prod_{i=1}^n\,
\int_{0}^{s_{i-1}} \mathrm{d}s_i~\sum_{{\mbf q}_i\in\zed^2}
\dot\xi^{\mu_i}(s_i)\,a_{{\mbf q}_i; \mu_i} \nonumber\\ && \qquad\times~
\e^{-\ii{\mbf q}_1\cdot({\mbf x} + \mbf\xi(s_1))/r'}
\star \cdots \star \e^{-\ii{\mbf q}_n\cdot({\mbf x} +
\mbf\xi(s_n))/r'}
\end{eqnarray}
which matches the $n^{\rm th}$ term in the expansion of
(\ref{NCwloop}).

It follows that the Morita correspondence between closed Wilson lines,
unlike the case of open Wilson
lines~\cite{Ambjorn:2000cs,Guralnik:2001pv}, does not involve any
transformation of the quantum numbers associated with the
loop. Thus if we take a closed Wilson line in the noncommutative
gauge theory which encloses an area $\rho$, then it maps into a closed
Wilson line in the Morita equivalent commutative gauge theory with
the \textit{same} shape and the \textit{same} area $\rho$, because
in the steps that led to (\ref{equivloops}) the parametrization
$\mbf\xi(s)$ of the loop never played any role. But, because of the relation
(\ref{transR2}), while the area of the loop remains fixed, the
area of the target torus is smaller. Thus the path can start
to wind in a non-trivial way and in general self-intersections of the
loop may appear in the dual gauge theory. While in the above analysis we
have focused only on the particular case when the noncommutativity parameter is
rational-valued, our conclusions concerning loop areas also hold in the more
general case when $\Theta$ is an irrational
number~\cite{Ambjorn:2000cs}.

\section{Dual Loop Correlators: Rational Case\label{DLCR}}

In this section we will explicitly compute some noncommutative
Wilson loop correlators. After choosing the closed path, we will apply
Morita duality to the observable (\ref{NCwloop}) thus mapping it into
a Wilson loop on a new torus. Since the target torus is
smaller, the dual Wilson loop can wind around the torus and
self-intersect in a very non-trivial way. To keep the discussion as
simple as possible and to provide concrete examples, we will restrict
ourselves to the case where the noncommutativity parameter is a
rational number and thus the target torus is a commutative space. We
will describe the case of irrational $\Theta$ in the next section.

As we will find, the actual computation of the Wilson loop average
depends heavily on the geometrical shape and orientation of
the closed path in the {\it original} torus. As the area of the
target torus is smaller than that of the original torus, under
Morita equivalence a simple closed curve can transform into a rather
intricate (self-intersecting) loop. In particular, it can happen that two
contractible loops, of different shape but equal area, transform
into two topologically inequivalent loops. The same is true of
contours which have the same area and shape but different
orientations on the torus. Paths which exhibit such behaviour can in
principle have very different quantum averages. In this section we
shall argue that this is indeed the case. We will focus on the
behaviour of a loop of fixed area on a shrinking torus. A complete
characterization of this phenomenon would take us deep into non-planar graph
theory, which is beyond the scope of this paper. Instead, we will
develop a working knowledge of this behaviour by
discussing a necessary criterion for a path to become
self-intersecting under Morita equivalence.

Consider a closed contractible path $\mathcal{C}$ with no
self-intersections on a square torus of radius $r$. Let
$\mbf\xi = (\xi^1, \xi^2) : [0,1]\rightarrow \mathbb{T}^2$ be a
parametrization of $\mathcal{C}$. Introduce the two {\it
  characteristic lengths}
\beq
\ell^\mu\left( \mathcal{C} \right) = \sup_{s,s' \in [0,1]}\,
\bigl|\xi^\mu(s)-\xi^\mu(s'\,)\bigr| \ , \quad \mu=1,2
\label{charlengths}\eeq
which measure the width and the height of the loop.
Given $a,b\in\zed$ and $\Theta\in\real$, consider the behaviour of the
path $\mathcal{C}$ as the torus shrinks to a torus of radius
\beq
r_{\rm c}^{a,b} = | a + b\,\Theta |\,r \ .
\label{rcritdef}\eeq
If $\ell^1 (\mathcal{C})$ and $\ell^2 (\mathcal{C})$ are
both smaller than $r_{\rm c}^{a,b}$, then the path will {\it not}
self-intersect on the dual torus. We thereby arrive at a necessary
condition that the loop $\mathcal{C}$ should satisfy in order to
self-intersect on the dual torus given by
\begin{equation} \label{bound}
\ell^\mu\left( \mathcal{C} \right) \ge r_{\rm c}^{a,b}
\end{equation}
for $\mu=1$ or $\mu=2$. We stress that the bound (\ref{bound}) is not
a sufficient condition. It is not difficult to draw loops that
do indeed satisfy the bound (\ref{bound}) but do not
self-intersect on the dual torus. In fact, a little practice with
drawing loops on the torus shows how involved the task
of providing necessary {\it and} sufficient conditions for
self-intersections is.

Through Morita equivalence, we can compute quantum averages of Wilson
loops on a noncommutative torus by mapping the observable to a
smaller but commutative torus and then resorting to the known
techniques of commutative Yang--Mills theory. But, according to
(\ref{bound}), given a loop on the original torus, there exists a {\it
critical radius} $r_{\rm c}^{a,b}$ such that the loop can become a
self-intersecting closed contour on the target torus. In the Morita
transformation to commutative gauge theory, the critical radius is
$r_{\rm c}^{a,b}=r/N$. We will now explore some of the physical
consequences of this statement.

\subsection{General Construction\label{GenConstr}}

In the previous section we have reduced the problem of evaluating a
noncommutative Wilson loop correlator to the computation of its
Morita dual correlator. We will now describe how this is
done in practice. According to~\cite{Witten:1991we,Migdal:1975zg,Rusakov1}, the
partition function of two-dimensional Yang--Mills theory on a torus
$\torus^2$ (and more generally on any Riemann surface) can be
conveniently evaluated through a combinatorial approach wherein one
covers the surface with a set of simplices (plaquettes) and works with
the lattice regularization of the original gauge theory . The
continuum limit is recovered in the limit as the
triangulation becomes finer. The partition function is
invariant under subdivision of the lattice, and thus the lattice
regularization provides a concrete definition of two
dimensional quantum Yang--Mills theory. In the lattice gauge theory,
the partition function is a sum over local factors associated to all
of the plaquettes, which each have the topology of a disk. It is
natural to associate to each plaquette $D_{\lambda}$ the holonomies
$U_{\sigma}$ of a gauge connection $A$ along its links
$L_{\sigma}$. Gauge invariance requires that the local factor
corresponding to each plaquette be a class function of the
holonomies.

The local factors $\Gamma (\mathcal{U}_\lambda;D_{\lambda})$
associated to each simplex $D_{\lambda}$ of area $\rho_{\lambda}$ are
given by~\cite{Migdal:1975zg}
\begin{equation} \label{migdal}
\Gamma(\mathcal{U}_\lambda;D_{\lambda}) = \sum_{R_{\lambda}}\,\dim R_{\lambda}~
\e^{- \frac{g^2\,\rho_{\lambda} }{2}\,C_2(R_{\lambda})}~
\chi^{~}_{R_{\lambda}} \left( \mathcal{U}_{\lambda} \right)
\end{equation}
where the sum runs through all isomorphism classes of $U(N)$
representations, $C_2(R_{\lambda})$ is the second Casimir invariant of the
representation $R_{\lambda}$, and $\chi^{~}_{R_{\lambda}}
(\mathcal{U}_{\lambda})=\mathrm{Tr}^{~}_{R_\lambda}\,\mathcal{U}_\lambda$ are
the
characters of the representation $R_{\lambda}$ evaluated on the holonomy
$\mathcal{U}_{\lambda}=\prod_\sigma U_\sigma$ along the
perimeter of the simplex $D_{\lambda}$ with respect to a fixed
orientation of its edges. The factors appearing in the formula
(\ref{migdal}) can be understood as follows. The representation
dimension $\dim R_{\lambda}$ is a normalization factor which
ensures that the holonomy around a loop of area $\rho_\lambda$
approaches $1$ as $\rho_\lambda\to 0$. The characters appear since they
form a basis for the vector space of class functions. Finally, the
exponential factor is essentially the exponential of the Yang--Mills
hamiltonian in the representation basis with $g^2$ the Yang--Mills
coupling constant. For a more detailed account see~\cite{Cordes:1994fc}.

Let us now consider the vacuum expectation value of a
Wilson loop in ordinary Yang--Mills theory. It is defined by the
functional integral
\begin{equation} \label{wloop}
W_{\mathcal{C} ; R} (\rho^{~}_{\mathcal{C}}) = \int
\DD A ~\e^{-S_{\rm YM}[A]}~\mathrm{Tr}^{~}_R~{\rm P}~\exp\left(\ii
\oint_{\mathcal{C}}A\right) \ ,
\end{equation}
where $\rho^{~}_{\mathcal{C}}$ is the area enclosed by the path
$\mathcal{C}$, $S_{\rm YM}[A]=\frac1{2g^2}\,\int_{\torus^2}\dd^2\mbf
x~\mathrm{Tr}_N^{~}\,F^2$ is the Yang--Mills action functional, and we have
explicitly indicated the dependence on the
representation $R$ of the character used to compute the holonomy of
the connection $A$ around the path $\mathcal{C}\subset\torus^2$. In
our case, the Wilson loop will always be taken to lie in the
fundamental representation $R=N$ of the $U(N)$ gauge group.

The Wilson loop provides a natural division of the torus into plaquettes
$D_{\lambda}$ bounded by line segments $L_{\sigma}$ (links in the
lattice formulation) in which the loop is divided by its
self-intersections. Each plaquette $D_\lambda$ has area $\rho_{\lambda}$
such that $\sum_{\lambda} \rho_{\lambda} = (2\pi\,r/N)^2$ is
the area of the Morita dual torus. In this way we can write
(\ref{wloop}) as
\begin{equation} \label{Wgeneral}
W_{\mathcal{C} ; R}(\rho^{~}_{\mathcal{C}})= \prod_\sigma\,\int_{U(N)}
\left[ \mathrm{d} U_{\sigma} \right]~
\prod_{\lambda}\,\Gamma \left(\mathcal{U}_\lambda;D_{\lambda} \right)
{}~\chi^{~}_{R}\left(\mathcal{U}^{-1}\right)
\end{equation}
where $\mathcal{U}=\prod_\sigma U_\sigma$ is the holonomy along the
corresponding edges $L_\sigma$ and $\left[ \mathrm{d} U_\sigma\right]$
is the invariant Haar measure on the $U(N)$ gauge group. In (\ref{Wgeneral}) we
have implicitly assumed that each plaquette has the topology of a
disk. If this is not the case, then it suffices to consider a
finer triangulation of the torus. For a simplex of different topology,
the local factor (\ref{migdal}) becomes~\cite{Cordes:1994fc}
\begin{equation} \label{migdalgeneral}
\Gamma(\mathcal{U}_\lambda;D_{\lambda}) = \sum_{R_{\lambda}} \left( \dim
R_{\lambda} \right)^{2-2h_\lambda-b_\lambda}~\e^{-
\frac{g^2\,\rho_{\lambda}}{2}
\,C_2(R_{\lambda})}~\chi^{~}_{R_{\lambda}} \left( \mathcal{U}_{\lambda}
\right)
\end{equation}
when the simplex $D_{\lambda}$ has $h_\lambda$ handles and $b_\lambda$
boundaries.

We can recast (\ref{Wgeneral}) in a simpler form by noticing that
each group element $U_{\sigma}$ representing the holonomy along
the edge $L_{\sigma}$ appears three times in the integral (once in
the Wilson line insertion and once for each of the two simplices that
has $L_{\sigma}$ as part of its boundary). Thus if we denote by
$R_{\alpha}(U)^{a}_{~b}$, $a,b=1,\dots,\dim R_\alpha$ the matrix
representing the group element $U$ in the representation $R_\alpha$,
then from the identity
\beq
\chi^{~}_{R_\alpha}\left(U\,U'\,\right) = R_{\alpha}
\bigl(U\bigr)^{a}_{~b}~R_{\alpha}
\left(U'\,\right)^{b}_{~a}
\label{repmatrices}\eeq
it follows that the computation of (\ref{Wgeneral}) reduces to
the evaluation of integrals of the form $\int_{U(N)}[\mathrm{d} U]~R_{\alpha}
(U)^{a}_{~b}\,R_{\beta} (U)^{c}_{~d}\,R_{\gamma} (U)^{e}_{~f}$.
Such group integrals give information about
the fusion numbers $\mathrm{N}_{~~R_{\alpha} R_{\beta}}^{R_{\gamma}}$ which
count the multiplicity of the irreducible representation
$R_{\gamma}$ in the Clebsch--Gordan decomposition $R_{\alpha}
\otimes R_{\beta} = \bigoplus_{R_{\gamma}}\mathrm{N}_{~~R_{\alpha}
R_{\beta}}^{R_{\gamma}}\,R_{\gamma}$.
We can collect these coefficients into factors associated with each
vertex of the triangulation which combine into a local
object. We may thereby write a final compact expression for the
Wilson loop average as
\begin{equation} \label{finalWloop}
W_{\mathcal{C} ; R}(\rho^{~}_{\mathcal{C}})= \sum_{R_{\lambda}}\,
\sum_{\varepsilon_{\sigma}}\,\prod_{\lambda}\,(\dim
R_{\lambda})^{2-2h_\lambda-b_\lambda}~\e^{- \frac{g^2\,\rho_{\lambda}}{2}\,
C_2(R_{\lambda})}~\prod_{\delta}\,\mathrm{G}_{\delta} \left( R,
R_{\lambda} ; \varepsilon_{\sigma} \right)
\end{equation}
where the index $\delta$ runs over all vertices of the
lattice, while $\varepsilon_{\sigma}$ runs over a basis for the
vector space of intertwiners between the representations
$R_{\gamma}$ and $R_{\alpha} \otimes R_{\beta}$. In the particular
case that the vertex $\delta$ is four-valent, the local factor
$\mathrm{G}_\delta$ is a $6j$-symbol
\cite{Witten:1991we,Cordes:1994fc}. We will see explicitly how this
works in some concrete examples below.

When the commutative gauge theory is related to noncommutative
Yang--Mills theory on $\torus_\theta^2$ by Morita duality, one uses the
global group isomorphism $U(N)=U(1)\times SU(N)/\zed_N$ to cancel the $U(1)$
contribution to the partition function by the background abelian gauge field
generated in the Morita transformation
(\ref{Moritagen})~\cite{GSV1}. One is then left with an $SU(N)/\zed_N$
gauge theory in a certain discrete
theta-vacuum~\cite{witten} of 't~Hooft flux $k=0,1,\dots,N$ which
labels the isomorphism classes of principal $SU(N)/\zed_N$ bundles
over the torus. The $U(1)$ phases only contribute non-trivially when
one sums over the topological sectors. For trivial bundles ($k=0$),
all formulas above hold using $SU(N)$ representations in place of
$U(N)$ representations. For non-trivial bundles ($k\neq0$), one
incorporates the background flux as follows. It contributes a factor
$\exp(\ii\oint_{\mathcal{C}}\alpha)$ to the Wilson loop
average~\cite{Ambjorn:2000cs}, where $\alpha$ is
any abelian gauge potential that gives rise to the constant background
flux $\Phi=\dd\alpha$. Then the dependence of
the correlator (\ref{wloop}) on the $k$ units of magnetic flux
follows from
\begin{equation} \label{thooftflux}
\frac1{2\pi}\,\oint_{\mathcal{C}}\alpha = \frac1{2\pi}\,
\int_{\Sigma}\Phi=\frac kN \ ,
\end{equation}
where $\Sigma=\partial\mathcal{C}$ is any surface spanned by the loop
$\mathcal{C}$. When $\mathcal{C}$ contains self-intersections, one has
to give a precise meaning to this integration.
The path $\mathcal{C}$ admits a unique decomposition
into simple closed paths as $\mathcal{C} =
\bigcup_i \mathcal{C}_i$. To obtain the appropriate flux factors one
then splits the holonomy line integral over $\mathcal{C}$ into line
integrals along the individual paths $\mathcal{C}_i$ and
repeatedly applies (\ref{thooftflux}). This modifies the characters in
the above formulas by products of the characters
$\chi_{R_\lambda}^{~}(\e^{2\pi\ii k/N})$ evaluated on elements in the
center of the $SU(N)$ gauge group.

\subsection{Simple Loops\label{Simple}}

We will now perform several explicit calculations in the rational
noncommutative gauge theory.
To illustrate the ideas in a somewhat general setting, we begin by comparing
the Wilson loop correlators associated to two paths which have the
generic forms depicted in
Figs.~\ref{figpath1} and~\ref{figpath2} (The torus $\torus^2$ is
throughout represented as a square of sides $r$ with opposite edges
identified). The paths enclose the same area
$\rho_1$, but the second one satisfies the inequality~(\ref{bound}).
\DOUBLEFIGURE{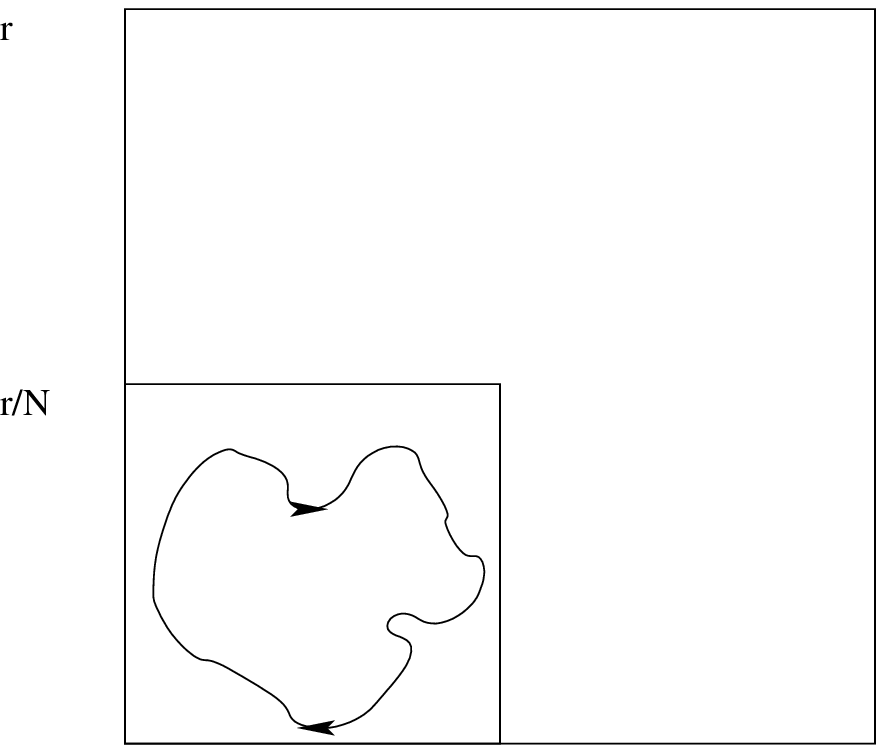,width=2.5in}{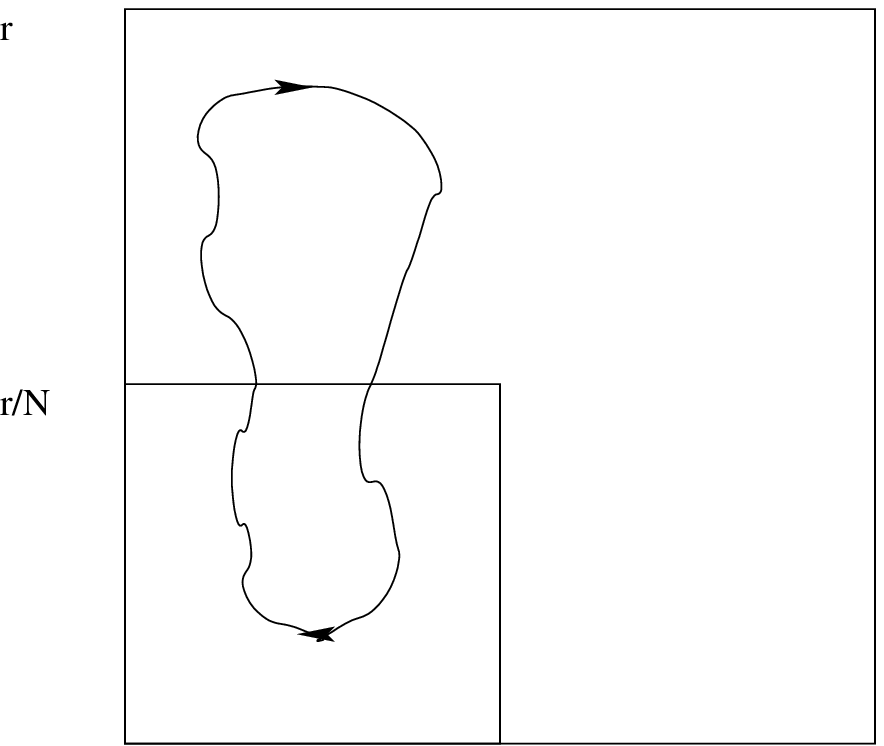,width=2.5in}{The
path $\mathcal{C}_1$. It is not affected by the shrinking of the torus.
\label{figpath1}}{The path $\mathcal{C}_2$. It contains non-trivial
self-intersections on the dual torus. \label{figpath2}}

Using (\ref{Wgeneral}) and (\ref{migdalgeneral}) the first loop
correlator can be associated with the formal expression
\bea \label{path1U(N)}
W_{\mathcal{C}_1;R}^k(\rho_1)&=&\sum_{R_1, R_2}\,\frac{ \dim
R_1} {\dim R_2}~\e^{- \frac{g^2\,\rho_1}{2}\,C_2(R_1)-
\frac{g^2\,\rho_2}{2}\,C_2(R_2)}~\chi^{~}_{R_1}\left(\e^{2 \pi\ii k/N}
\right)\nonumber\\ && \qquad \times~\int_{SU(N)}[
\mathrm{d} U_1 ]~\chi^{~}_{R_1} (U_1)~\chi^{~}_{R_2}\left(U_1^{-1}\right)
{}~\chi_{R}^{~} (U_1) \nonumber\\ &=& \sum_{R_1, R_2}\,\frac{ \dim
R_1} {\dim R_2}~\mathrm{N}_{~~R_1 R}^{R_2}~
\e^{- \frac{g^2\,\rho_1}{2}\,C_2(R_1)-
\frac{g^2\,\rho_2}{2}\,C_2(R_2)}~\chi^{~}_{R_1}\left(\e^{2 \pi\ii k/N}
\right)
\end{eqnarray}
with $\rho_1+\rho_2=(2\pi\,r/N)^2$, where we have performed the group
integrals to obtain the fusion
coefficients of the three representations. The irreducible
representations $R$ of $SU(N)$ can be labelled by decreasing sets
${\mbf n}^R=(n_1^R,\dots,n_N^R)$ of $N$ integers, $+ \infty > n^R_1 >
n^R_2 > \cdots > n^R_N > - \infty$, which satisfy the linear Casimir
constraint $\sum_{a=1}^Nn_a^R=0$. They determine the lengths of the
rows of the corresponding Young tableaux. In particular, the integer
$\sum_{a=1}^{N-1}n_a^R$ is the total number of boxes in the Young diagram
describing $R$. In terms of these integers,
the second Casimir invariant of $R$ can be written as
\begin{equation}
C_2 (R) = C_2 \left( {\mbf n}^R\right) = \sum_{a=1}^N \left( n_a^R -
\frac{N-1}{2}
\right)^2 - \frac{N}{12}\,\left(N^2 -1\right)+\frac{\left(n_N^R
\right)^2}N \ ,
\end{equation}
while the dimension of $R$ can be expressed as the Vandermonde
determinant
\begin{equation}
\dim R = \Delta\left(\mbf n^R\right)=\prod_{a<b}\left(n_a^R-n_b^R
\right) \ .
\label{dimRVan}\end{equation}
To compute the fusion numbers, we use the Weyl formula for the
$SU(N)$ characters
\begin{equation}
\chi^{~}_R (U) = \chi^{~}_{{\mbf n}^R} \left(\e^{2 \pi\ii \mbf\lambda}
\right)= \frac{\det\limits_{1 \le a,b \le N}\,\left[\e^{2 \pi \ii n^R_a\,
      \lambda_b}\right]}{\Delta \left(\e^{2 \pi\ii\mbf\lambda} \right)}
\end{equation}
where $\e^{2 \pi \ii
  \mbf\lambda}=(\e^{2\pi\ii\lambda_1},\dots,\e^{2\pi\ii\lambda_N})$,
$\lambda_a \in [0,1]$, $a=1,\dots,N$ are the eigenvalues of the
unitary matrix $U$ with $\sum_{a=1}^N\lambda_a=0~\mathrm{mod}~\zed$. Then the
integration over the group variables $U$ can be
transformed into an integration over the eigenvalues at the price
of introducing a jacobian $\Delta(\e^{2 \pi \ii \mbf\lambda})^2$. With
these identifications, we can finally write (\ref{path1U(N)}) for
$R=N$ the fundamental representation as
\begin{eqnarray} \label{path1U(N)final}
W_{\mathcal{C}_1;N}^k(\rho_1)&=&\sum_{{\mbf n}^{R_1},
{\mbf n}^{R_2}}\,\frac{ \Delta\left(\mbf n^{R_1}\right)}
{\Delta\left(\mbf n^{R_2}\right)}~\e^{- \frac{g^2\,
\rho_1}{2}\,C_2({\mbf n}^{R_1})-
\frac{g^2\,\rho_2}{2}\,C_2( {\mbf n}^{R_2})}~\e^{\frac{2
\pi\ii k}{N}\,\sum\limits_{a=1}^{N-1}n_a^{R_1}}
\nonumber\\ && \qquad \times\,\prod_{a=1}^N\,\int_0^1 \mathrm{d} \lambda_a~
\delta\left(\,\mbox{$\sum\limits_{a=1}^N$}\,\lambda_a\right)~\sum_{c=1}^N
\e^{2 \pi \ii \lambda_c}\nonumber \\ &&\qquad\times\,
\det_{1 \le a,b \le N} \left[\e^{2
\pi \ii n^{R_1}_a\,\lambda_b} \right]\,\det_{1 \le a,b \le N}
\left[\e^{2 \pi \ii n^{R_2}_a\,\lambda_b} \right] \ .
\end{eqnarray}
We will return to the evaluation of this expression in
Section~\ref{Wilsoninf}.

\begin{figure}
\begin{center}
\includegraphics[width=2.5in]{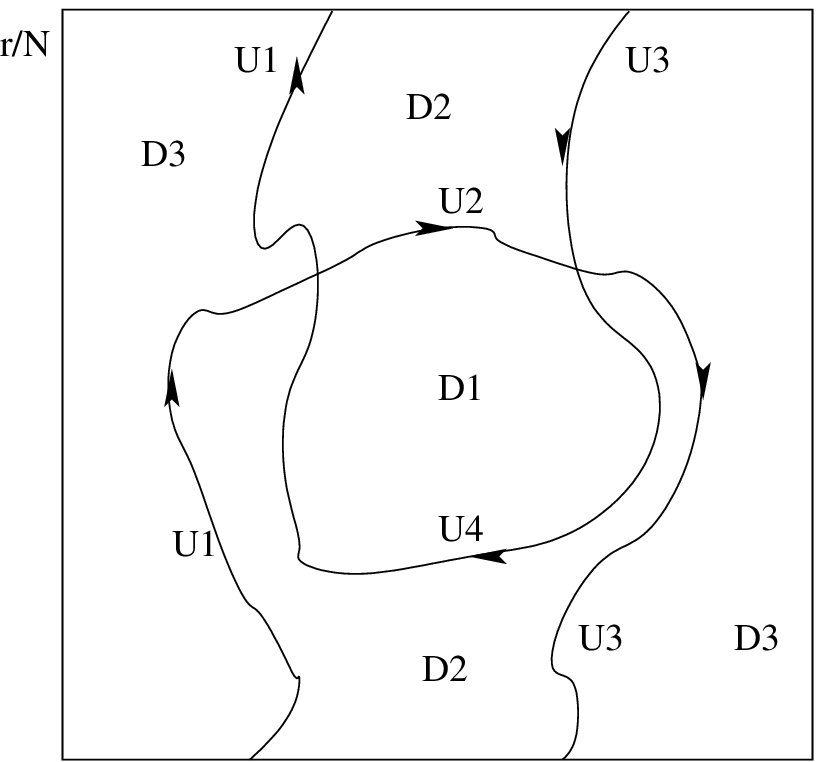}
\caption{On the dual torus the loop $\mathcal{C}_2$ looks like a
  non-trivial self-intersecting path.} \label{figpath2big}
\end{center}
\end{figure}

The calculation is much different for the second path.
Let us associate to the domains depicted in Fig.~\ref{figpath2big}
the local factors (\ref{migdalgeneral}) given by
\begin{eqnarray}
&& \Gamma \left(\mathcal{U}_1; D_1 \right) =  \sum_{R_1}\,
\dim R_1~\e^{-\frac{g^2\,
\rho'_1}{2}\,C_2(R_1)}~\chi_{R_1}^{~} (U_2\,U_4) \ , \nonumber \\ &&
\Gamma \left(\mathcal{U}_2; D_2
\right) = \sum_{R_2}\,\dim R_2~\e^{-\frac{g^2\,\rho'_2}{2}\,C_2(R_2)}~
\chi^{~}_{R_2}\left(U_1\,U_4^{-1}\,U_3\,U_2^{-1}\right) \ , \nonumber \\ &&
\Gamma \left(\mathcal{U}_3; D_3
\right) = \sum_{R_3}\,\frac1{\dim R_3}~\e^{-\frac{g^2\,\rho'_3}{2}\,C_2(R_3)}~
\chi^{~}_{R_3}\left(U_1^{-1}\right)~\chi^{~}_{R_1}\left(U_3^{-1}\right)
\ ,
\end{eqnarray}
where the dual area parameters obey
$\rho_1'+\rho_2'+\rho_3'=(2\pi\,r/N)^2$ and
$2\rho_1'+\rho_2'=\rho_1$. The last factor can be understood by
regarding the contribution from the third simplex as a cylinder
amplitude whose initial and final states are
parametrized by the holonomies $U_1$ and $U_3$.
Then the general formula (\ref{Wgeneral}) for the path $\mathcal{C}_2$
becomes
\begin{eqnarray} \label{path2}
W_{\mathcal{C}_2;R}^k(\rho_1) & = & \sum_{R_1,R_2,R_3} \frac{
\dim R_1\,\dim R_2} {\dim R_3}~\e^{-\frac{g^2\,\rho'_1}{2}\,
C_2(R_1) -\frac{g^2\,\rho'_2}{2}\,C_2(R_2)-\frac{g^2\,\rho'_3}{2}\,
C_2(R_3)}~ \nonumber\\ && \times\,
\prod_{\sigma=1}^4\,\int_{SU(N)}[ \mathrm{d} U_\sigma]~\chi^{~}_{R_1}
(U_2\,U_4)~\chi^{~}_{R_2}\left(U_1\,U_4^{-1}\,U_3\,U_2^{-1}
\right)~\chi^{~}_{R_3}
\left(U_1^{-1}\right)~\chi^{~}_{R_3}\left(U_3^{-1}\right) \nonumber\\ &&
\times\,
\chi^{~}_{R} (U_1\,U_2\,U_3\,U_4)~\chi_{R_1}^{~}\left(\e^ {2 \pi\ii k/N }
\right)^2~
\chi^{~}_{R_2} \left(\e^ {2 \pi\ii k/N} \right) \ ,
\end{eqnarray}
where the 't Hooft flux factors arise from the
decomposition of the holonomy integral
\begin{equation}
\oint_{\mathcal{C}_2}\alpha=
\sum_{\sigma=1}^4\,\int_{L_\sigma}\alpha = \left(\,\oint_{L_1\cup L_4^{-1}\cup
    L_3\cup L_2^{-1}} + 2\,\oint_{L_4\cup L_2}\,\right)\alpha
\end{equation}
and the line segment $L_\sigma$ refers to the path labelled by the
holonomy $U_\sigma$ in Fig.~\ref{figpath2big}. Thus one of the central
characters squares in (\ref{path2}). Employing the same $SU(N)$
representation machinery used to arrive at (\ref{path1U(N)final}), we
can rewrite (\ref{path2}) as
\begin{eqnarray}\label{path2U(N)final}
W_{\mathcal{C}_2;R}^k(\rho_1)& = & \sum_{{\mbf n}^{R_1},
{\mbf n}^{R_2}, {\mbf n}^{R_3}}\,\frac{ \Delta
\left(\mbf n^{R_1}\right)\,\Delta\left(\mbf n^{R_2}\right)}
{\Delta\left(\mbf n^{R_3}\right)}~\e^{
-\frac{g^2\,\rho'_1}{2}\,C_2( {\mbf n}^{R_1})
-\frac{g^2\,\rho'_2}{2}\,C_2( {\mbf n}^{R_2})
-\frac{g^2\,\rho_3'}{2}\,C_2( {\mbf n}^{R_3})}
\nonumber \\ && \times\,\prod_{\sigma=1}^4\,\int_{SU(N)}
[ \mathrm{d} U_\sigma]~\chi^{~}_{R_1}
(U_2\,U_4)~\chi^{~}_{R_2}\left(U_1\,U_4^{-1}\,U_3\,U_2^{-1}
\right)~\chi^{~}_{R_3}
\left(U_1^{-1}\right)~\chi^{~}_{R_3}\left(U_3^{-1}\right) \nonumber \\
&& \times\,\chi^{~}_{R} (U_1\,U_2\,U_3\,U_4)~\e^{\frac{2\pi\ii
  k}{N}\,\sum\limits_{a=1}^{N-1}(2n_a^{R_1}+n_a^{R_2})} \ .
\end{eqnarray}

This is as far as we can proceed with general expressions for the
Wilson loop correlators (\ref{path1U(N)final}) and
(\ref{path2U(N)final}). Superficially, these two analytic expressions
look quite different. For example, while (\ref{path1U(N)final})
depends only on the loop area $\rho_1$, the function
(\ref{path2U(N)final}) effectively depends on {\it two} independent
areas, say $\rho_1$ and $\rho_1'$. If the dependence on $\rho_1'$ is
non-trivial, then evidently the two loop correlators are distinct,
even though in the original theory they enclosed the same area
$\rho_1$. To perform a more direct comparison of these correlation
functions and get an idea of the nature of this extra area dependence,
let us simplify matters enormously by turning to the special example
of $SU(2)$ gauge theory. In this case we may appeal to various
well-known angular momentum identities from the representation
theory of the group $SU(2)$~\cite{Varshalovich:1988ye}.

Irreducible representations $R_j$ of $SU(2)$ are labelled by an
angular momentum quantum number $j\in\frac12\,\nat_0$.
The dimension of $R_j$ is given by $\dim R_j = 2 j + 1$, the quadratic
Casimir invariant is $C_2 (R_j) = j\,(j + 1)$, and the total number of
boxes in the Young diagram representing $R_j$ is $2j$. The
integrations over the group variables in (\ref{path1U(N)}) give the
fusion numbers $\mathrm{N}^{j_2}_{~~j_1 j}$ which count the multiplicity
of the irreducible representations $R_{j_2}$ in the Clebsch--Gordan
decomposition of $R_{j_1} \otimes R_{j}$. These
coefficients are equal to~$1$ if $|j_1 - j| \le j_2 \le j_1 +
j$ and~$0$ otherwise. Thus we can write the quantum average
(\ref{path1U(N)}) for $SU(2)$ gauge group in the explicit form
\beq \label{path1explicit}
W_{\mathcal{C}_1;j}^k(\rho_1)=\sum_{j_1\in\frac12\,\nat_0}~
\sum_{j_2=|j_1 - j|}^{j_1 + j}\,(-1)^{2 j_1\,k}~
\mbox{$\frac {2 j_1 +1 } {2 j_2 + 1}$}~\e^{- \frac{g^2\,
\rho_1}{2}\,j_1\,(j_1 + 1)- \frac{g^2\,
\rho_2}{2}\,j_2\,(j_2 + 1)} \ .
\eeq

Now let us rewrite the expression (\ref{path2}) by using the explicit
form of the characters for $SU(2)$ representations. In this
case we can introduce as representation matrices the Wigner
functions of angular momentum $j$, so that
(\ref{repmatrices}) becomes
\begin{equation}
\chi^{~}_{R_j}\left(U\,U'\,\right) = {\sf D}^{j}_{~mm'}
\bigl(U\bigr)~{\sf D}^{j}_{~m'm}\left(U'\,\right)
\end{equation}
with $-j\leq m,m'\leq j$, where throughout we implicitly
assume that repeated indices
represented by lower case Latin letters are summed over.
In this way we can better organize the integration over $SU(2)$ group
variables and write (\ref{path2}) as
\begin{eqnarray} \label{genpath1}
W_{\mathcal{C}_2;j}^k(\rho_1)&=&\sum_{j_1,j_2,j_3\in
\frac12\,\nat_0}\,(-1)^{2j_2\,k}~
\frac{(2j_1+1)\,(2j_2+1)} {2j_3+1} \nonumber\\ && \qquad\times~
\e^{-\frac{g^2\,\rho'_1}{2}\,
j_1\,(j_1+1) -\frac{g^2\,\rho'_2 }{2}\,j_2\,(j_2+1) -\frac{g^2\,\rho'_3}{2}
\,j_3\,(j_3+1)} \nonumber\\ && \qquad\times~\int_{SU(2)}
[ \mathrm{d} U_1 ]~{\sf D}^{j_2}_{~~a_2
b_2}(U_1)~{\sf D}^{j_3}_{~~a_3 a_3}\left(U_1^{-1}\right)~
{\sf D}^{j}_{~a b}(U_1)\nonumber\\ && \qquad\times~
\int_{SU(2)} [ \mathrm{d} U_2 ]~{\sf D}^{j_1}_{~~a_1 b_1}(U_2)~
{\sf D}^{j_2}_{~~d_2 a_2}\left(U_2^{-1}\right)~
{\sf D}^{j}_{~b c}(U_2) \nonumber\\ && \qquad\times~\int_{SU(2)}
[ \mathrm{d} U_3 ]~{\sf D}^{j_2}_{~~c_2 d_2}(U_3)~{\sf D}^{j_3}_{~~b_3
b_3}\left(U_3^{-1}\right)~{\sf D}^{j}_{~c d}(U_3)\nonumber\\ && \qquad\times~
\int_{SU(2)} [ \mathrm{d} U_4
]~{\sf D}^{j_1}_{~~b_1 a_1}(U_4)~{\sf D}^{j_2}_{~~b_2 c_2}\left(U_4^{-1}\right)
{}~{\sf D}^{j}_{~d a}(U_4) \ .
\end{eqnarray}
If we regard the path drawn in
Fig.~\ref{figpath2big} as a triangulation of the torus as before, then each
Wigner function ${\sf D}^{j}_{~mm'}(U)$ is associated with
an oriented edge of the triangulation, with the first index $m$
representing the origin of the line segment and the second index $m'$
representing its endpoint. The reason for this
identification is that it is more convenient to understand the
quantum average (\ref{genpath1}) as a product of contributions arising from the
\textit{vertices} of the triangulation, rather than as integrals over
edge variables. This procedure implements the general construction of
Section~\ref{GenConstr} and provides an explicit realization of the
expression~(\ref{finalWloop}).

To this end we use the formula~\cite{Varshalovich:1988ye}
\begin{equation} \label{integralU}
\Bigl[{}^{j_1}_{m_1}~{}^{j_2}_{m_2}~{}^{j_3}_{m_3} \Bigr]\,
\Bigl[{}^{j_1}_{m_1'}~{}^{j_2}_{m'_2}~{}^{j_3}_{m'_3} \Bigr]=
\frac{ 2 j_3 + 1 }{8 \pi^2}\,\int_{SU(2)}[\mathrm{d} U]~{\sf D}^{j_1}_{~~m_1
m'_1}(U)~{\sf D}^{j_2}_{~~m_2 m'_2}(U)~\overline{{\sf D}^{j_3}_{~~m_3 m'_3}(U)}
\end{equation}
relating the integral over edge variables to a product of two Clebsch--Gordan
coefficients, each one associated with an endpoint of the given
edge. We can now perform the integration over
group variables in (\ref{genpath1}) and collect together the
Clebsch--Gordan coefficients associated to each vertex, which in the
present case are all of valence~$4$. The crucial
identity is~\cite{Varshalovich:1988ye}
\begin{eqnarray} \label{CGto6j}
&& \sum_{m_1,m_2,m_3,m_{12},m_{23}}\,\Bigl[{}^{j_{12}}_{m_{12}}~{}^{
    j_3}_{m_3}~{}^{j}_{ m}\Bigr]\,\Bigl[{}^{j_1}_{ m_1}~{}^{ j_2}_{
    m_2}~{}^{j_{12}}_{ m_{12}}\Bigr]\,\Bigl[{}^{j_1}_{ m_1}~
{}^{ j_{23}}_{m_{23}}~{}^{j'}_{ m'}\Bigr]\,\Bigl[{}^{j_2}_{ m_2}~{}^{
  j_3}_{ m_3}~{}^{j_{23}}_{ m_{23}}\Bigr] \nonumber\\ &&
\qquad\qquad~=~\delta_{j
j'}~\delta_{m m'}~(-1)^{j_1 + j_2 + j_3 + j}~\sqrt{(2 j_{12} +
1)\,(2 j_{23} +
1)}~\left\{{}^{j_1}_{j_3}~{}^{j_2}_j~{}^{j_{12}}_{j_{23}}\right\} \ .
\end{eqnarray}
We have introduced the classical Wigner $6j$-symbol whose explicit
form is provided by the Racah formula. We will not require this
detailed expression here, except for noting that the square of the
$6j$-symbol in (\ref{CGto6j}) is proportional to a product of
completely symmetric combinatorial factors
$\triangle(j_1,j_2,j_{12})\,\triangle(j_1,j,j_{23})\,
\triangle(j_3,j_2,j_{23})\,\triangle(j_3,j,j_{12})$ which are each
non-vanishing only if the triangle inequality
\beq
\triangle(j_1,j_2,j_3)~:~ \quad j_1\leq j_2+j_3 \ ,
\quad j_2\leq j_1+j_3 \ , \quad j_3\leq j_1+j_2 \ ,
\quad j_1+j_2+j_3\in\nat_0+\mbox{$\frac12$}
\label{triangleins}\eeq
is obeyed by the corresponding angular momenta. In computing
(\ref{genpath1}) the triangle inequalities imply that the average is
non-zero only for half-integer spin $j$, in which case the loop
correlator is given explicitly by
\bea \label{genpath2}
W_{\mathcal{C}_2;j}^k(\rho_1)&=&\sum_{2j_2\geq j+\frac12}~
\prod_{{\alpha<\beta}\atop{\alpha\neq2}}~
\sum^{j_\beta+j}_{{j_\alpha=|j_\beta-j|}
\atop{j\leq j_\alpha+j_\beta\in\nat_0}}\,(-1)^{2
  j_2\,k}~\mbox{$\frac {(2 j_1 + 1)\,(2 j_2 + 1)} {2 j_3 +
1}$}~\left\{{}^j_j~{}^{j_1}_{j_2}~{}^{j_2}_{j_3}
\right\}^2\nonumber\\ && \qquad\times~\e^{-\frac{g^2\,\rho'_1}{2}\,
j_1\,(j_1+1) -\frac{g^2\,\rho'_2 }{2}\,j_2\,(j_2+1) -\frac{g^2\,\rho'_3}{2}
\,j_3\,(j_3+1)}
\eea
up to an irrelevant overall numerical factor.

The key point now is that the expressions (\ref{path1explicit}) and
(\ref{genpath2}), while bearing certain similarities, are very
different. In particular, the correlator
(\ref{genpath2}) appears to be a non-trivial function of the extra
area $\rho_1'$. After writing the areas of the second and third
simplices as $\rho_2'=\rho_1-2\rho_1'$ and $\rho_3'=\rho_2+\rho_1'$, we
can differentiate (\ref{genpath2}) with respect to $\rho_1'$ to get
\begin{eqnarray} \label{areadep}
\frac{\partial W_{\mathcal{C}_2;j}^k(\rho_1)}{\partial\rho_1'}
&=&-\frac{g^2}2~\sum_{2j_2\geq j+\frac12}~
\prod_{{\alpha<\beta}\atop{\alpha\neq2}}~
\sum^{j_\beta+j}_{{j_\alpha=|j_\beta-j|}
\atop{j\leq j_\alpha+j_\beta\in\nat_0}}\,(-1)^{2
  j_2\,k}~\left\{{}^j_j~{}^{j_1}_{j_2}~{}^{j_2}_{j_3}\right\}^2\nonumber
\\ && \qquad\times~\e^{-\frac{g^2\,\rho_1}{2}\,j_2\,(j_2+1)
-\frac{g^2\,\rho_2}{2}
\,j_3\,(j_3+1)}~\e^{-\frac{g^2\,\rho_1'}2\,[j_1\,(j_1+1)
-2j_2\,(j_2+1)+j_3\,(j_3+1)]}\nonumber\\ && \qquad\times~
\mbox{$\frac {(2 j_1 + 1)\,(2 j_2 + 1)\,\bigl[j_1\,(j_1+1)-
2j_2\,(j_2+1)+j_3\,(j_3+1)\bigr]} {2 j_3 +1}$} \ .
\end{eqnarray}
We have not been able to rigorously prove that this quantity is
non-vanishing. But we have also not been able to find any angular
momentum identities implying that it is~$0$, and we strongly doubt the
existence of any such identity. Asymptotically, while the
$6j$-symbol has an exponential decay for certain configurations of large
angular momenta~\cite{Roberts1}, generically it has only a
trigonometric behaviour for large $j$'s. When $\rho_1'\gg0$, one can
construct an area-preserving diffeomorphism on the noncommutative
torus which changes $\rho_1'$ and therefore likely gives a different
correlator on the commutative torus. Thus the convergent series
(\ref{areadep}) does not appear to produce a vanishing result. This
heuristic argument is strong evidence in favour of
the non-vanishing of the expression (\ref{areadep}). We therefore
propose that for the class of simple loops considered here, the
corresponding Wilson averages are strongly dependent on the shapes and
even the orientations of the contours on $\torus^2_\theta$.

\subsection{Circular Loops\label{Circle}}

To explore further the shape and orientation dependence of rational
noncommutative Wilson loops, let us now consider a more specific
smooth path with the circular geometry of Fig.~\ref{circle}. Under
Morita equivalence it is mapped to the complicated self-intersecting
path of Fig.~\ref{circlebig}.
\DOUBLEFIGURE{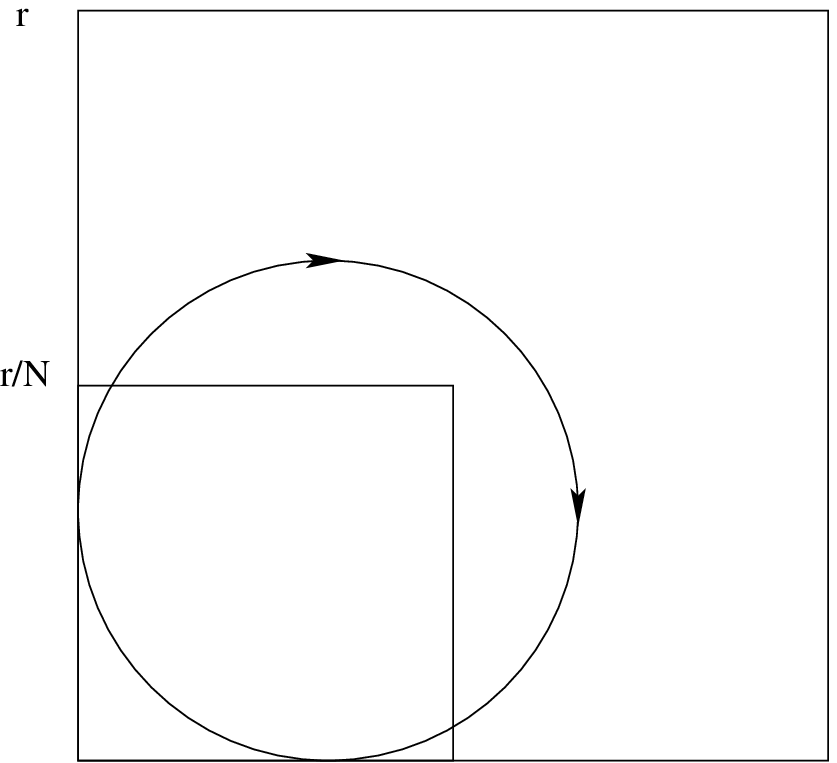,width=2.5in}{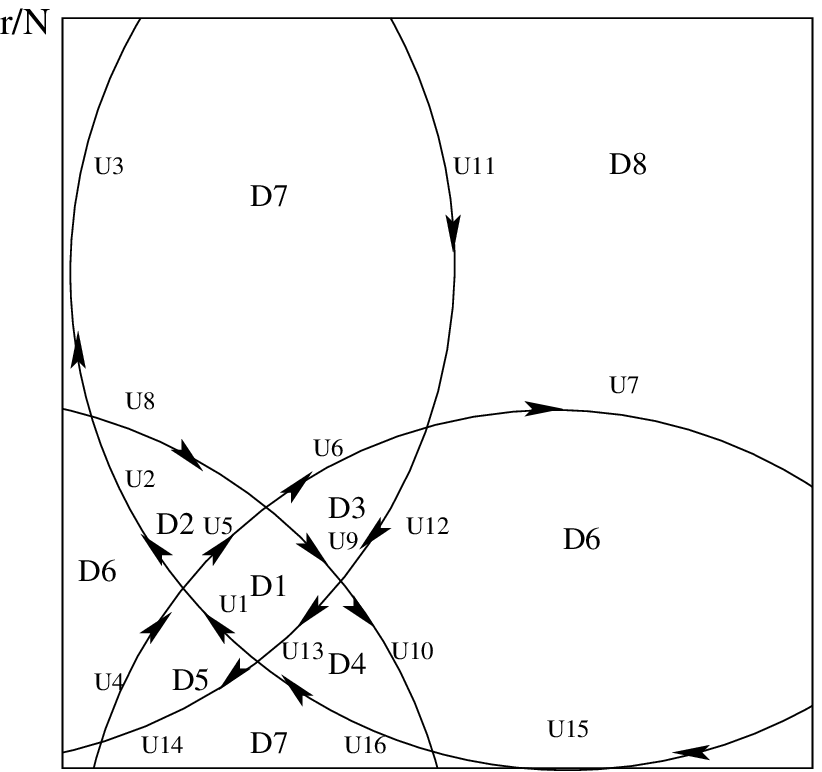,width=2.5in}{The
  circular loop on the original torus.
\label{circle}}{The self-intersecting path on the target torus into
which the circular loop is mapped under the Morita
  transformation. \label{circlebig}}
Using the general formula
(\ref{Wgeneral}), we must associate to each simplex $D_\lambda$, which
in this case all have the topology of a disk, the local
factor (\ref{migdal}). We label the simplices and the corresponding
edges, each with their proper orientation, as shown in
Fig.~\ref{circlebig}. We take the Wilson loop in the representation
$R$. With this notation, the circular Wilson loop correlator then
reads
\begin{eqnarray}
W_{\bigcirc;R}^k(\rho_1)&=&\sum_{{R_1},\dots,R_8}~\prod_{\lambda=1}^{8}\,\dim
R_\lambda~
\e^{-\frac{g^2\,\rho'_\lambda}{2}\,C_2(R_\lambda)}\,
\chi_{R_1}^{~}\left(\e^{2\pi \ii k/N} \right)^3\,  \chi_{R_2}^{~}
\left(\e^{2\pi \ii k/N} \right)^2
\nonumber\\ &&\times\,\chi_{R_3}^{~}\left(\e^{2\pi \ii k/N} \right)^2\,
\chi_{R_4}^{~}\left(\e^{2\pi \ii k/N} \right)^2\,\chi^{~}_{R_5}\left(
\e^{2\pi \ii k/N} \right)^2\,\chi_{R_6}^{~}\left(\e^{2\pi\ii k/N}
\right)\nonumber\\ &&
\times\,\chi^{~}_{R_7} \left(\e^{2\pi \ii k/N} \right)\,
\prod_{\sigma=1}^{16}\,\int_{SU(N)}[ \mathrm{d}U_\sigma]~
\chi_{R_1}^{~}\left( U_{1}\, U_{5}\, U_{9}\, U_{14} \right)
\,\chi_{R_2}^{~} \left( U_{2}\, U_{8}\, U_{5}^{-1} \right)
\nonumber\\ &&\times\,\chi_{R_3}^{~}
\left( U_{6}\, U_{12}\, U_{9}^{-1} \right)\,\chi_{R_4}^{~} \left( U_{16}\,
U_{13}^{-1}\, U_{10} \right)\,\chi_{R_5}^{~} \left( U_{14}\, U_{4}\,
U_{1}^{-1} \right)\nonumber\\ &&\times\,\chi_{R_6}^{~} \left( U_{15}\,
U_{10}^{-1}
\,U_{12}^{-1}\, U_{7}\, U_{2}^{-1}\, U_{4}^{-1} \right)\,\chi_{R_7}^{~} \left(
U_{8}^{-1}\, U_{3}\, U_{14}^{-1}\, U_{16}^{-1}\, U_{11}\, U_{6}^{-1} \right)
\nonumber\\ && \times\,\chi_{R_8}^{~} \left( U_{7}^{-1}\, U_{11}^{-1}\,
U_{15}^{-1}
\,U_{3}^{-1} \right)\,\chi_{R}^{~}
\left(\,\mbox{$\prod\limits_{\sigma=1}^{16}\,U_\sigma$}\right)
\end{eqnarray}
where we have used
\begin{equation}
\oint_\bigcirc\alpha=\sum_{\sigma=1}^{16}\,\int_{L_{\sigma}}\alpha
=\left(3\,\oint_{\partial D_1} +
2\,\oint_{\partial D_2} + 2\,\oint_{\partial D_3} + 2\,
\oint_{\partial D_4} + 2\, \oint_{\partial D_5} + \oint_{\partial
D_6} + \oint_{\partial D_7}\right)\alpha
\end{equation}
and the dual areas $\rho_\lambda'$ obey
\beq
\sum_{\lambda=1}^8\,\rho_\lambda'=\left(\frac{2\pi\,r}N\right)^2 \ ,
\quad 4\rho_1'+3\rho_2'+3\rho_3'+3\rho_4'+3\rho_5'+2\rho_6'+2\rho_7'
+\rho_8'=\rho_1 \ .
\label{circleareas}\eeq

As above, in order to be as explicit as possible we will limit
the analysis to the case of an $SU(2)$ gauge group.
We write the characters in terms of Wigner functions and integrate
over each group variable individually using (\ref{integralU}).
Then we collect together all Clebsch--Gordan coefficients
relative to each vertex, which are again all of valence~$4$, to get
\begin{eqnarray}
&& W_{\bigcirc;j}^k(\rho_1)=\sum_{j_1,\dots,j_8\in\frac12\,\nat_0}~
\mbox{$\frac{(-1)^{2(j_1+j_6+j_7)\,k}}
{\bigl[(2 j_1 + 1)\,(2 j_6 + 1)\,(2 j_7 + 1)\,(2 j_8 +1)\bigr]^3}$}
{}~\e^{-\sum\limits_{\lambda=1}^8\frac{g^2\,\rho_\lambda'}
{2}\,j_\lambda\,(j_\lambda+1)}
\nonumber\\ && \qquad\times\,\left(\Bigl[{}^{j_1}_{ b_1}~{}^{ j}_b~{}^{j_5}_{
    c_5}\Bigr]\,\Bigl[{}^{j_1}_{ b_1}~{}^j_e~{}^{j_2}_{a_2}\Bigr]\,
\Bigl[{}^{j_5}_{ c_5}~{}^j_e~{}^{j_6}_{ f_6}\Bigr]\,
\Bigl[{}^{j_2}_{ a_2}~{}^j_b~{}^{j_6}_{ f_6}\Bigr]\right)\,
\left(\Bigl[{}^{j_1}_{ c_1}~{}^j_f~{}^{j_2}_{ c_2}\Bigr]\,
\Bigl[{}^{j_2}_{c_2}~{}^j_i~{}^{j_7}_{ a_2}\Bigr]\,
\Bigl[{}^{j_3}_{ a_3}~{}^j_f~{}^{j_7}_{ a_7}\Bigr]\,
\Bigl[{}^{j_1}_{ c_1}~{}^j_i~{}^{j_3}_{ a_3}\Bigr]\right)
\nonumber\\ &&\qquad\times\,\left(\Bigl[{}^{j_1}_{ d_1}~{}^j_q~{}^{j_3}_{
    c_3}\Bigr]\,\Bigl[{}^{j_3}_{ c_3}~{}^j_m~{}^{j_6}_{c_6}\Bigr]\,
\Bigl[{}^{j_4}_{ c_4}~{}^j_q~{}^{j_6}_{ c_6}\Bigr]\,
\Bigl[{}^{j_1}_{ d_1}~{}^j_m~{}^{j_4}_{ f_4}\Bigr]\right)\,
\left(\Bigl[{}^{j_1}_{ a_1}~{}^j_n~{}^{j_4}_{ b_4}\Bigr]\,
\Bigl[{}^{j_4}_{b_4}~{}^j_a~{}^{j_7}_{ d_7}\Bigr]\,
\Bigl[{}^{j_5}_{ a_5}~{}^j_n~{}^{j_7}_{ d_7}\Bigr]\,
\Bigl[{}^{j_1}_{ a_1}~{}^j_a~{}^{j_5}_{ a_5}\Bigr]\right)
\nonumber\\ &&\qquad\times\,\left(\Bigl[{}^{j_2}_{ b_2}~{}^j_c~{}^{j_6}_{
    e_6}\Bigr]\,\Bigl[{}^{j_6}_{ e_6}~{}^j_h~{}^{j_8}_{a_8}\Bigr]\,
\Bigl[{}^{j_7}_{ b_7}~{}^j_c~{}^{j_8}_{ a_8}\Bigr]\,
\Bigl[{}^{j_2}_{ a_2}~{}^j_h~{}^{j_7}_{ b_7}\Bigr]\right)\,
\left(\Bigl[{}^{j_3}_{ b_3}~{}^j_g~{}^{j_7}_{ f_7}\Bigr]\,
\Bigl[{}^{j_7}_{f_7}~{}^j_l~{}^{j_8}_{ b_8}\Bigr]\,
\Bigl[{}^{j_6}_{ d_6}~{}^j_g~{}^{j_8}_{ b_8}\Bigr]\,
\Bigl[{}^{j_3}_{ b_3}~{}^j_l~{}^{j_6}_{ d_6}\Bigr]\right)
\nonumber\\ &&\qquad\times\,\left(\Bigl[{}^{j_4}_{
    a_4}~{}^j_r~{}^{j_6}_{b_6}\Bigr]\,
\Bigl[{}^{j_6}_{ b_6}~{}^j_p~{}^{j_8}_{c_8}\Bigr]\,
\Bigl[{}^{j_4}_{ a_4}~{}^j_p~{}^{j_7}_{ e_7}\Bigr]\,
\Bigl[{}^{j_7}_{ e_7}~{}^j_k~{}^{j_8}_{ c_8}\Bigr]\right)\,
\left(\Bigl[{}^{j_5}_{ b_5}~{}^j_o~{}^{j_7}_{ c_7}\Bigr]\,
\Bigl[{}^{j_7}_{c_7}~{}^j_d~{}^{j_8}_{ b_8}\Bigr]\,
\Bigl[{}^{j_6}_{ a_6}~{}^j_o~{}^{j_8}_{ d_8}\Bigr]\,
\Bigl[{}^{j_5}_{ b_5}~{}^j_d~{}^{j_6}_{ a_6}\Bigr]\right)\nonumber\\ &&
\end{eqnarray}
where except for the $j$'s all Latin indices are implicitly summed
over. Each term in parentheses is the local representation of a
self-intersection on the Morita dual circle. It can be written in
a more compact way by repeatedly applying the formula (\ref{CGto6j}), as
well as various symmetry properties of the Clebsch--Gordan
coefficients that can be found in~\cite{Varshalovich:1988ye}, to
convert four Clebsch--Gordan coefficients into a $6j$-symbol. Up to
an overall numerical factor one finally finds
\begin{eqnarray} \label{circlefinal}
W_{\bigcirc;j}^k(\rho_1)&=&\sum_{j_1,\dots,j_8\in D_\bigcirc^j}\,
(-1)^{2(j_1+j_6+j_7)\,k}\,\prod_{\lambda=1}^{8}\,(2 j_{\lambda} + 1)~
\e^{- \frac{g^2\,\rho'_{\lambda}}{2}\, j_{\lambda}\,( j_{\lambda} +
  1)}\nonumber\\ && \qquad
\times\,\left\{{}^j_j~{}^{j_1}_{j_6}~{}^{j_2}_{j_5}\right\}\,
\left\{{}^j_j~{}^{j_1}_{j_7}~{}^{j_3}_{j_2}\right\}\,
\left\{{}^j_j~{}^{j_1}_{j_6}~{}^{j_4}_{j_3}\right\}\,
\left\{{}^j_j~{}^{j_1}_{j_7}~{}^{j_5}_{j_4}\right\}
\nonumber\\ && \qquad\times\,
\left\{{}^j_j~{}^{j_2}_{j_8}~{}^{j_7}_{j_6}\right\}\,
\left\{{}^j_j~{}^{j_3}_{j_8}~{}^{j_6}_{j_7}\right\}\,
\left\{{}^j_j~{}^{j_4}_{j_8}~{}^{j_6}_{j_7}\right\}\,
\left\{{}^j_j~{}^{j_5}_{j_8}~{}^{j_6}_{j_7}\right\} \ ,
\end{eqnarray}
where by the triangle inequalities the sum over spins is restricted to
the range
\beq
D_\bigcirc^j=\bigcup_{{\alpha=2,3,4,5}\atop{\beta=1,6,7}}\,
D_{~j_\alpha j_\beta}^j~~\cup~~\bigcup_{\alpha=6,7}\,
D_{~j_8j_\alpha}^j
\label{circrange}\eeq
with
\beq
D^j_{~j_\alpha j_\beta}=\left\{|j_\beta-j|\leq j_\alpha\leq
  j_\beta+j~,~j_\alpha+j_\beta\geq j~,~
j+j_\alpha+j_\beta\in\nat_0+\mbox{$\frac12$}\right\} \ .
\label{Djrangedef}\eeq
In contrast to the intersecting Wilson loop average over the contour
$\mathcal{C}_2$ of Section~\ref{Simple}, the correlator
(\ref{circlefinal}) is generically non-vanishing for all angular
momenta $j\in\frac12\,\nat_0$.

\subsection{Square Loops\label{Square}}

For our final explicit example, we will consider the case of the
polygonal contour with the geometry of the square Wilson loop of
Fig.~\ref{square}.
\DOUBLEFIGURE{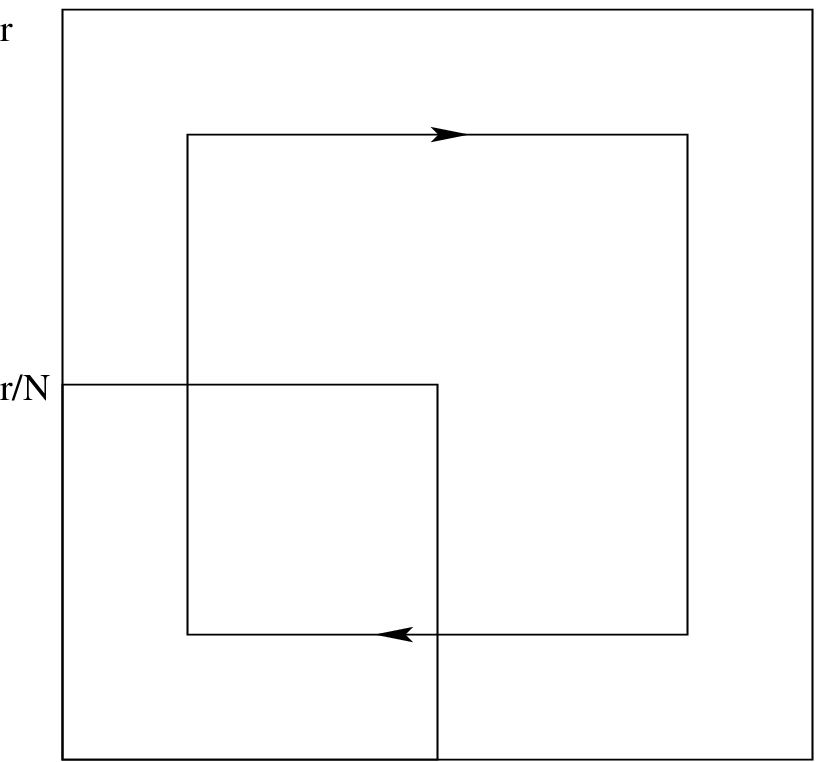,width=2.5in}{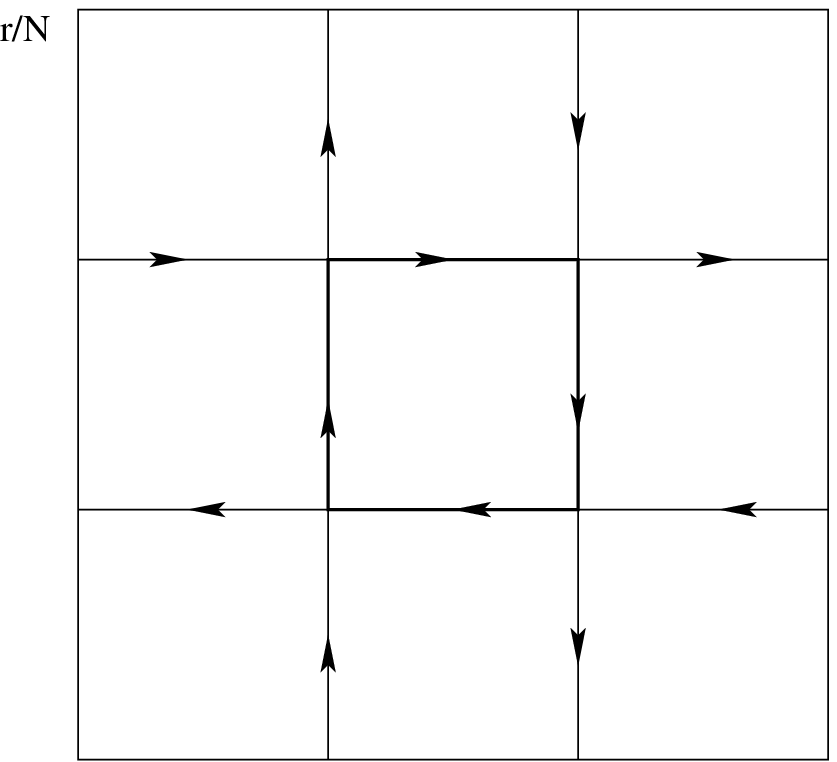,width=2.5in}{The
square loop on the original torus. \label{square}}{The square loop on the
target torus. The edges that bound the inner square are covered
twice by the loop. \label{squareself}}
After a Morita transformation this loop is mapped into the loop of
Fig.~\ref{squareself}. This dual path is much more complicated than
the previously considered dual circle, because the edges which bound
the inner square of Fig.~\ref{squareself} are covered twice in
computing the Wilson loop holonomy using the combinatorial
construction of Section~\ref{GenConstr}. Thus the group elements
associated with these particular edges will appear four times in
(\ref{Wgeneral}), twice because of the Wilson loop holonomy and
once for each of the two faces that are bounded by this
edge due to (\ref{migdal}). We would then need the
generalization of the group integral (\ref{integralU}) involving four
group elements, but these generalizations are difficult to handle.
Because of this technical difficulty, instead of computing the square
Wilson loop of Fig.~\ref{square}, we will perform an area-preserving
deformation of the square contour as illustrated in Fig.~\ref{defsquare}.
After a Morita transformation, this path is mapped to the loop
of Fig.~\ref{defsquareself}. Each simplex $D_\lambda$ in this
case has the topology of a disk.
\DOUBLEFIGURE{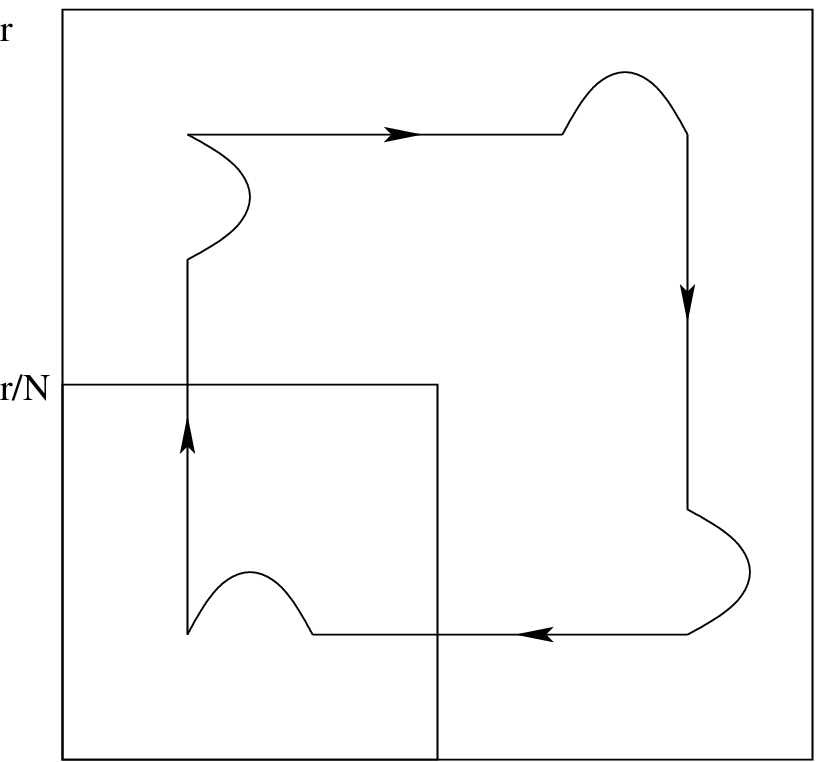,width=2.5in}{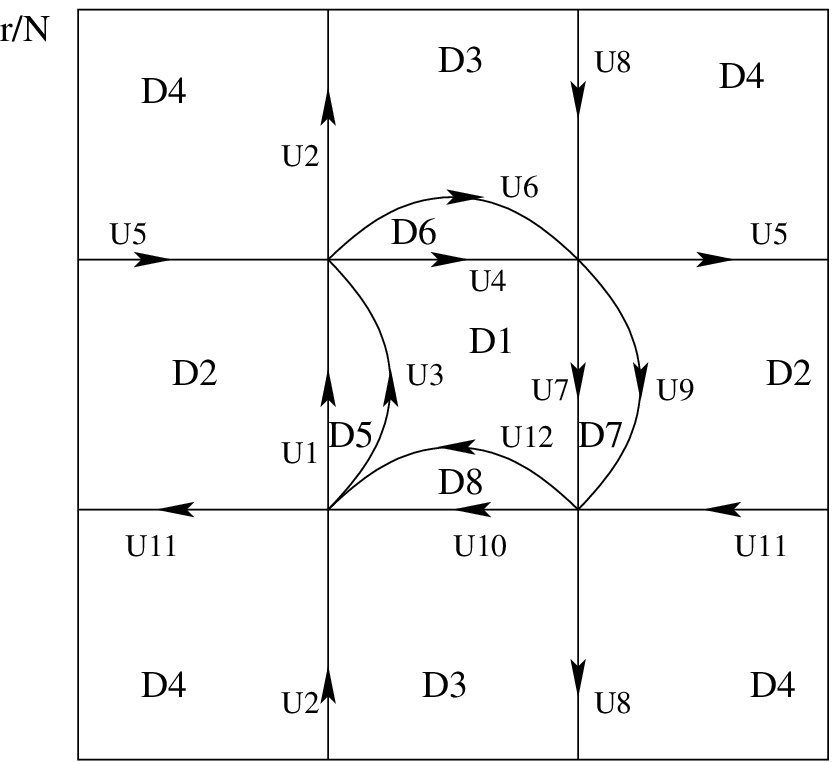,width=2.5in}{The
deformed square loop on the original torus. \label{defsquare}}{The
deformed square loop on the target torus. No edge is covered
more than once. \label{defsquareself}}
The condition that the loop of
Fig.~\ref{defsquare} encloses the same area as the loop of
Fig.~\ref{square} implies for the areas $\rho_\lambda'$ of the simplices
$D_\lambda$ of Fig.~\ref{defsquareself} that
\begin{equation} \label{areacondition}
\rho'_6 + \rho'_7 = \rho'_5 + \rho'_8 \ .
\end{equation}

The deformed square Wilson loop correlator (\ref{Wgeneral}) in the
commutative dual gauge theory thereby reads
\begin{eqnarray} \label{squarestep1}
W^k_{\Box;R}(\rho_1)&=&\sum_{{R_1},\dots,R_8}~\prod_{\lambda=1}^{8}\,\dim
R_\lambda~
\e^{-\frac{g^2\,\rho'_\lambda}{2}\,C_2(R_\lambda)}\,
\chi_{R_1}^{~}\left(\e^{2\pi \ii k/N} \right)^3\,  \chi_{R_2}^{~}
\left(\e^{2\pi \ii k/N} \right)
\nonumber\\ &&\times\,\chi_{R_3}^{~}\left(\e^{2\pi \ii k/N} \right)\,
\chi_{R_5}^{~}\left(\e^{2\pi \ii k/N} \right)^2\,\chi^{~}_{R_6}\left(
\e^{2\pi \ii k/N} \right)^2\,\chi_{R_7}^{~}\left(\e^{2\pi\ii k/N}
\right)^2\nonumber\\ &&
\times\,\chi^{~}_{R_8} \left(\e^{2\pi \ii k/N} \right)^2\,
\prod_{\sigma=1}^{12}\,\int_{SU(N)}[ \mathrm{d}U_\sigma]~
\chi_{R_1}^{~}\left( U_{3} \,U_{4}\, U_{7}\, U_{12}  \right)
\,\chi_{R_2}^{~} \left(U_{11} \,U_{9}^{-1}\, U_{5}\, U_{1}^{-1}  \right)
\nonumber\\ &&\times\,\chi_{R_3}^{~}
\left( U_{2}\, U_{10}^{-1}\, U_{8}\, U_{6}^{-1}\right)\,\chi_{R_4}^{~}
\left(U_{8}^{-1}\, U_{11}^{-1}\, U_{2}^{-1}\, U_{5}^{-1}
\right)\,\chi_{R_5}^{~}\left( U_{1} \,U_{3}^{-1}\right)
\nonumber\\ &&\times\,\chi_{R_6}^{~} \left(U_{6}\,
U_{4}^{-1}\right)\,\chi_{R_7}^{~} \left(U_{9}\, U_{7}^{-1} \right)\,
\chi_{R_8}^{~} \left( U_{10}\, U_{12}^{-1} \right)\,\chi_{R}^{~}
\left(\,\mbox{$\prod\limits_{\sigma=1}^{12}\,U_\sigma$}\right)
\end{eqnarray}
where we have used
\begin{equation}
\oint_\Box\alpha=\sum_{\sigma=1}^{12}\,\int_{L_{\sigma}}\alpha=
\left(3\,\oint_{\partial D_1} +
2\,\oint_{\partial D_5} + 2\,\oint_{\partial D_6} + 2\,
\oint_{\partial D_7} + 2\,\oint_{\partial D_8} + \oint_{\partial
D_2} + \oint_{\partial D_3}\right)\alpha
\end{equation}
and the dual areas $\rho_\lambda'$ obey, in addition to
(\ref{areacondition}), the constraints
\beq
\sum_{\lambda=1}^8\,\rho_\lambda'=\left(\frac{2\pi\,r}N\right)^2 \ ,
\quad 3\rho_1'+2\,\sum_{\lambda=2}^8\,\rho_\lambda'=\rho_1 \ .
\label{squareareaconstr}\eeq
Again we take the gauge group to be $SU(2)$ and follow our
combinatorial procedure. Each group integration is performed by using
the formula (\ref{integralU}). Each edge contributes to
(\ref{squarestep1}) with a Clebsch--Gordan coefficient for each one of
its endpoints. The sum over edges (holonomies) is converted into a sum over
vertices (collections of Clebsch--Gordan coefficients). However,
now each vertex of the triangulation depicted in Fig.~\ref{defsquareself}
is of valence~$6$ and so will generally have associated to it a
more complicated object than a $6j$-symbol. By collecting the
Clebsch--Gordan coefficients for each of the four vertices, the
quantum average (\ref{squarestep1}) becomes
\begin{eqnarray} \label{squarestep2}
W^k_{\Box;j}(\rho_1)&=&\sum_{j_1,\dots,j_8\in\frac12\,\nat_0}~
\mbox{$\frac{(-1)^{2(j_1+j_2+j_3)\,k}~(2j_1+1)}
{(2 j_2 + 1)\,(2 j_3 + 1)\,(2 j_4 + 1)^3}$}
{}~\e^{-\sum\limits_{\lambda=1}^8\frac{g^2\,\rho_\lambda'}{2}\,
j_\lambda\,(j_\lambda+1)}
\nonumber\\ && \qquad\times\,\left(\Bigl[{}^{j_5}_{ a_5}~{}^j_a~{}^{j_2}_{
    a_2}\Bigr]\,\Bigl[{}^{j_1}_{ a_1}~{}^j_c~{}^{j_5}_{a_5}\Bigr]\,
\Bigl[{}^{j_1}_{ a_1}~{}^j_a~{}^{j_8}_{ b_8}\Bigr]\,
\Bigl[{}^{j_8}_{ b_8}~{}^j_b~{}^{j_3}_{ b_3}\Bigr]\,
\Bigl[{}^{j_3}_{ b_3}~{}^j_c~{}^{j_4}_{ c_4}\Bigr]\,
\Bigl[{}^{j_2}_{ a_2}~{}^j_k~{}^{j_4}_{ c_4}\Bigr]\right)_{\rm A}
\nonumber\\ && \qquad\times\,\left(\Bigl[{}^{j_5}_{ b_5}~{}^j_b~{}^{j_2}_{
    d_2}\Bigr]\,\Bigl[{}^{j_2}_{ d_2}~{}^j_f~{}^{j_4}_{d_4}\Bigr]\,
\Bigl[{}^{j_3}_{ a_3}~{}^j_b~{}^{j_4}_{ d_4}\Bigr]\,
\Bigl[{}^ {j_6}_{a_6}~{}^j_f~{}^{j_3}_{ a_3}\Bigr]\,
\Bigl[{}^{j_1}_{ b_1}~{}^j_d~{}^{j_6}_{ a_6}\Bigr]
\Bigl[{}^{j_1}_{ b_1}~{}^j_d~{}^{j_5}_{ b_5}\Bigr]\right)_{\rm B}
\nonumber\\ && \qquad\times\,\left(\Bigl[{}^{j_1}_{ c_1}~{}^j_g~{}^{j_7}_{
    a_7}\Bigr]\,\Bigl[{}^{j_1}_{ c_1}~{}^j_e~{}^{j_6}_{b_5}\Bigr]\,
\Bigl[{}^{j_6}_{ b_6}~{}^j_g~{}^{j_3}_{ d_3}\Bigr]\,
\Bigl[{}^{j_3}_{ d_3}~{}^j_i~{}^{j_4}_{ a_4}\Bigr]\,
\Bigl[{}^{j_2}_{ c_2}~{}^j_e~{}^{j_4}_{ c_4}\Bigr]\,
\Bigl[{}^{j_1}_{ c_1}~{}^j_g~{}^{j_2}_{c_2}\Bigr]\right)_{\rm C}
\nonumber\\ && \qquad\times\,\left(\Bigl[{}^{j_1}_{ d_1}~{}^j_h~{}^{j_7}_{
    b_7}\Bigr]\,\Bigl[{}^{j_7}_{ b_7}~{}^j_q~{}^{j_2}_{b_2}\Bigr]\,
\Bigl[{}^{j_2}_{ b_2}~{}^j_l~{}^{j_4}_{ b_4}\Bigr]\,
\Bigl[{}^{j_3}_{ c_3}~{}^j_h~{}^{j_4}_{ b_4}\Bigr]\,
\Bigl[{}^{j_8}_{ a_8}~{}^j_q~{}^{j_3}_{ c_3}\Bigr]\,
\Bigl[{}^{j_1}_{ d_1}~{}^j_l~{}^{j_8}_{ a_8}\Bigr]\right)_{\rm D}
\end{eqnarray}
where for later reference we have labelled each vertex
contribution with an upper case Latin letter.

Let us now consider in more detail the individual vertex contributions
in (\ref{squarestep2}). Their computation relies on a number of
angular momentum identities which can all be found
in~\cite{Varshalovich:1988ye}. We begin with the vertex labelled
`A'. The first three Clebsch--Gordan coefficients can be summed by
using the formula
\begin{equation} \label{formulaCG1}
\sum_{\alpha, \beta ,\delta}\,\Bigl[{}^{a}_{
  \alpha}~{}^b_\beta~{}^{c}_\gamma
\Bigr]\,\Bigl[{}^{d}_\delta~{}^b_\beta~{}^{e}_\epsilon\Bigr]\,
\Bigl[{}^a_\alpha~{}^f_\phi~{}^d_\delta\Bigr]=
(-1)^{b+c+d+f}~\sqrt{(2 c + 1)\,(2 d +
  1)}~\Bigl[{}^c_\gamma~{}^f_\phi~{}^{e}_\epsilon\Bigr]~\Bigl\{
{}^a_e~{}^b_f~{}^c_d\Bigr\} \ ,
\end{equation}
while the last three coefficients can be summed in a similar way
thanks to the identity
\begin{equation} \label{formulaCG2}
\sum_{\alpha ,\beta
  ,\delta}\,\Bigl[{}^b_\beta~{}^c_\gamma~{}^a_\alpha\Bigr]\,
\Bigl[{}^b_\beta~{}^e_\epsilon~{}^d_\delta\Bigr]\,
\Bigl[{}^a_\alpha~{}^f_\phi~{}^d_\delta\Bigr]=
(-1)^{a+b+e+f}~\sqrt{\mbox{$\frac{2 a + 1}{2 e + 1}$}}~(2 d + 1)~
\Bigl[{}^c_\gamma~{}^f_\phi~{}^e_\epsilon\Bigr]~\Bigl\{
{}^a_e~{}^b_f~{}^c_d\Bigr\} \ .
\end{equation}
The first three Clebsch--Gordan coefficients of the vertex labelled `B'
can be summed similarly by again applying (\ref{formulaCG2}), while
the remaining Clebsch--Gordan contributions sum to Kronecker
delta-functions according to the orthogonality relations
\begin{eqnarray}
\sum_{\alpha, \beta}\,\Bigl[{}^a_\alpha~{}^b_\beta~{}^c_\gamma\Bigr]\,
\Bigl[{}^a_\alpha~{}^b_\beta~{}^{c'}_{\gamma'}\Bigr]&=&
\delta_{c c'}~\delta_{\gamma
\gamma'} \ , \nonumber\\ \sum_{\alpha, \gamma}\,
\Bigl[{}^a_\alpha~{}^b_\beta~{}^c_\gamma\Bigr]\,
\Bigl[{}^a_\alpha~{}^{b'}_{\beta'}~{}^c_\gamma\Bigr]&=&
\mbox{$\frac{2 c +1}{2 b +1}$}
{}~\delta_{b b'}~\delta_{\beta \beta'} \ .
\end{eqnarray}
The vertex C has the same structure as vertex A.

The final vertex D has a completely different structure.
By means of the reflection identity
\begin{equation}
\Bigl[{}^a_\alpha~{}^b_\beta~{}^c_\gamma\Bigr]= (-1)^{a-\alpha}~\sqrt{
\mbox{$\frac{2 c
+ 1}{2 b +1}$}}~\Bigl[{}^c_\gamma~{}^a_{-\alpha}~{}^b_\beta\Bigr]
\end{equation}
its contribution can be rewritten as
\bea
&&(-1)^{j_7 - j_2 + j_8 - j_3}~\mbox{$\frac{\sqrt{(2 j_7 + 1)\,(2 j_8 +1)}}
{2 j+ 1}$}\nonumber\\ && \qquad\qquad\times~
\Bigl[{}^j_q~{}^{j_7}_{b_7}~{}^{j_2}_{ b_2}\Bigr]\,
\Bigl[{}^{j_8}_{ a_8}~{}^{ j_1}_{ -d_1}~{}^j_l\Bigr]\,
\Bigl[{}^{j_2}_{ b_2}~{}^j_l~{}^{j_4}_{ b_4}\Bigr]\,
\Bigl[{}^j_q~{}^{ j_8}_{ a_8}~{}^{j_3}_{ c_3}\Bigr]\,
\Bigl[{}^{j_7}_{ b_7}~{}^{ j_1}_{ -d_1}~{}^j_h\Bigr]\,
\Bigl[{}^{j_3}_{ c_3}~{}^j_h~{}^{j_4}_{ b_4}\Bigr] \ .
\eea
We can then apply the identity
\begin{eqnarray}
&&
\sum_{\alpha,\beta,\dots,\nu}~\Bigl[{}^a_\alpha~{}^b_\beta~{}^c_\gamma\Bigr]\,
\Bigl[{}^d_\delta~{}^e_\epsilon~{}^f_\phi\Bigr]\,
\Bigl[{}^c_\gamma~{}^f_\phi~{}^q_\nu\Bigr]\,
\Bigl[{}^a_\alpha~{}^d_\delta~{}^g_\eta\Bigr]\,
\Bigl[{}^b_\beta~{}^e_\epsilon~{}^h_\mu\Bigr]\,
\Bigl[{}^g_\eta~{}^h_\mu~{}^q_\nu\Bigr] \nonumber\\ && \qquad\qquad =~
\sqrt{(2 c + 1)\,(2 f + 1)\,(2 g + 1)\,(2 h + 1)}~(2 q + 1)~
\left\{\begin{matrix}\scriptstyle a&~\scriptstyle d&~\scriptstyle g\\[0.1mm]
\scriptstyle b&~\scriptstyle e&~\scriptstyle h\\[0.1mm]
\scriptstyle c&~\scriptstyle f&~\scriptstyle q\end{matrix}\right\}
\end{eqnarray}
to obtain a final expression for the contribution from vertex D in
terms of $9j$-symbols of the second kind.

By grouping together all of these contributions, the deformed
square Wilson loop correlator (\ref{squarestep2}) thus becomes
\begin{eqnarray} \label{squarestep3}
W^k_{\Box;j}(\rho_1)&=&\sum_{j_1,\dots,j_8\in\frac12\,\nat_0}\,
(-1)^{2(j_1+j_2+j_3)\,k}~\delta_{j_5 j_6}~
\delta_{j_6 j_5}~ \delta_{a_5 b_6}~ \delta_{b_6 a_5}
{}~\e^{-\sum\limits_{\lambda=1}^8\frac{g^2\,\rho_\lambda'}{2}\,
j_\lambda\,(j_\lambda+1)}
\nonumber\\&&\qquad\times~(2j_1+1)\,(2j_4+1)\,(2j_7+1)\,(2j_8+1)~
\sqrt{\mbox{$\frac{2j_5+1}{2j_6+1}$}}\nonumber\\ &&\qquad\times~
\Bigl[{}^{j_8}_{ b_8}~{}^j_c~{}^{j_2}_{ a_2}\Bigr]\,
\Bigl[{}^{j_8}_{ b_8}~{}^j_c~{}^{j_2}_{a_2}\Bigr]\,
\Bigl[{}^{j_7}_{ a_7}~{}^j_e~{}^{j_3}_{ d_3}\Bigr]\,
\Bigl[{}^{j_7}_{ a_7}~{}^j_e~{}^{j_3}_{ d_3}\Bigr]\,
\biggl\{{}^{\stackrel{\scriptstyle j}{\scriptstyle j_8}}_{j_3}~
{}^{\stackrel{\scriptstyle j_7}{\scriptstyle j_1}}_j~
{}^{\stackrel{\scriptstyle j_2}{\scriptstyle j}}_{j_4}\biggr\}
\nonumber\\ && \qquad\times~
\left\{{}^{j_1}_{j_2}~{}^j_j~{}^{j_8}_{j_5}\right\}\,
\left\{{}^{j_3}_{j_2}~{}^j_j~{}^{j_8}_{j_4}\right\}\,
\left\{{}^{j_2}_{j_3}~{}^j_j~{}^{j_5}_{j_4}\right\}\,
\left\{{}^{j_2}_{j_3}~{}^j_j~{}^{j_7}_{j_4}\right\}\,
\left\{{}^{j_1}_{j_3}~{}^j_j~{}^{j_7}_{j_6}\right\} \ .
\end{eqnarray}
We can rewrite this expression in a manner which resembles more
closely the circular Wilson loop correlator (\ref{circlefinal}) by
expressing the $9j$-symbol in terms of
$6j$-symbols, at the price of having to introduce an additional
angular momentum sum. This is accomplished via the identity
\beq
\biggl\{{}^{\stackrel{\scriptstyle j}{\scriptstyle j_8}}_{j_3}~
{}^{\stackrel{\scriptstyle j_7}{\scriptstyle j_1}}_j~
{}^{\stackrel{\scriptstyle j_2}{\scriptstyle j}}_{j_4}\biggr\}
=\sum_{j_9\in\frac12\,\nat_0}(-1)^{3j+j_1+j_2+j_3+j_4+j_7+j_8+2j_9}\,
(2j_9+1)\,\left\{{}^j_j~{}^{j_3}_{j_7}
{}~{}^{j_9}_{j_8}\right\}\,\left\{{}^{j_7}_{j_4}~{}^j_{j_2}~{}^{j_9}_{j_1}
\right\}\,\left\{{}^{j_2}_j~{}^{j_4}_{j_3}~{}^{j_9}_j\right\} \ .
\label{9j6jid}\eeq
Doing the implicit sums left over in (\ref{squarestep3}) then gives
the final form
\bea
W^k_{\Box;j}(\rho_1)&=&\sum_{2j_4=0}^{4j-1}~\sum_{j_1,\dots,j_9\in
D_\Box^j}\,(-1)^{(j_1+j_2+j_3)\,(2k+1)+3j+j_4+j_7+j_8+2j_9}~
\prod_{{\alpha=1}\atop{\alpha\neq6}}^9(2j_\alpha+1)\nonumber\\ &&
\qquad\times ~\e^{-\frac{g^2\,(
\rho_5'+\rho_6')}2\,j_5\,(j_5+1)}~
\prod_{{\lambda=1}\atop{\lambda\neq5}}^8
\e^{-\frac{g^2\,\rho_\lambda'}{2}\,j_\lambda\,(j_\lambda+1)}
\nonumber\\ &&\qquad\times~
\left\{{}^{j_1}_{j_2}~{}^j_j~{}^{j_8}_{j_5}\right\}\,
\left\{{}^{j_3}_{j_2}~{}^j_j~{}^{j_8}_{j_4}\right\}\,
\left\{{}^{j_2}_{j_3}~{}^j_j~{}^{j_5}_{j_4}\right\}\,
\left\{{}^{j_2}_{j_3}~{}^j_j~{}^{j_7}_{j_4}\right\}\nonumber\\ &&
\qquad\times~
\left\{{}^{j_1}_{j_3}~{}^j_j~{}^{j_7}_{j_6}\right\}\,
\left\{{}^j_j~{}^{j_3}_{j_7}
{}~{}^{j_9}_{j_8}\right\}\,\left\{{}^{j_7}_{j_4}~{}^j_{j_2}~{}^{j_9}_{j_1}
\right\}\,\left\{{}^{j_2}_j~{}^{j_4}_{j_3}~{}^{j_9}_j\right\}
\label{squarefinal}\eea
where by the triangle inequalities the sum over spins is restricted to
the range
\beq
D_\Box^j=D^{j_1}_{~~j_2j_7}~~\cup~~D^{j_2}_{~~j_4j_9}~~
\cup~~D^j_{~j_1j_6}~~\cup~~D^j_{~j_2j_3}~~\cup~~
\bigcup_{{\alpha=4,5,7,8}\atop{\beta=1,2,3}}D^j_{~j_\alpha j_\beta}
{}~~\cup~~\bigcup_{\alpha=3,8}D^j_{~j_9j_\alpha} \ .
\label{squarerange}\eeq

We can now compare (\ref{squarefinal}) with the circular Wilson loop
correlator (\ref{circlefinal}) that encloses the same area $\rho_1$ as
the square on the original torus. Again, while bearing some
similarities, the two formulas have a very different angular momentum
structure and a different functional dependence on the areas
involved. It is thus very likely that they are
different. Of course this is not a rigorous proof that the two
expressions obtained are really not equal, and to accomplish this one
should perform the sum over all angular momenta. Unfortunately it
is very difficult to handle these sums analytically.

These calculations can be straightforwardly generalized to more
complicated polygonal contours on $\torus^2_\theta$. The differences
will lie in the nature of the corresponding triangulation of the dual
torus. The generic contribution from a local vertex will involve a
$3nj$-symbol of the second kind, which can be represented as a sum
over products of $n$ $6j$-symbols~\cite{Varshalovich:1988ye}. The
higher the valencies of these vertices the more angular momentum sums that
are introduced, yielding apparently distinct expressions for the corresponding
loop correlators. This is evident even in the additional area
dependences that the self-intersecting contours contain. While in
principle the dual areas $\rho_\lambda'$ depend on the original area
$\rho_1$ and the rank $N$ of the Morita dual commutative gauge theory,
an infinitesimal total area-preserving variation of the parameters
$\rho_\lambda'$ generally produces a non-vanishing result and accounts
for the distinct correlation functions obtained.

The claimed shape dependence of Wilson loops on the noncommutative torus is
much more drastic than on the noncommutative
plane~\cite{ADM1,BDPTV1}. For example, it is clear that a circular
contour and an ellipsoidal contour can produce distinct loop
correlators, even though the two loops can be mapped into one another
by a unimodular linear transformation. This can be understood from the
fact that the {\it global} $U(\infty)$ group of area-preserving
diffeomorphisms on $\torus^2$ is different from that on
$\real^2$~\cite{LSZ1}. Because of the smaller invariance
group on $\torus^2$, rotational symmetry is lost. Thus the loop
correlators depend crucially on the orientation in the torus and other
geometrical factors in addition to the shape of the contour. A similar
feature has been observed numerically in the lattice regularization of
the noncommutative gauge theory~\cite{BHN1}.
Within the present combinatorial approach,
the shape dependence of closed Wilson line correlators is understood
through an intricate graph theoretic problem. Note that, conversely,
an intricate self-intersecting Wilson loop described by a graph in
commutative non-abelian gauge theory can be mapped to a
simple Wilson loop in $U(1)$ noncommutative gauge
theory. The self-intersections can be thought of as being absorbed
into the noncommutativity of spacetime, in much the same way that the
rank $N$ can.

\section{Dual Loop Correlators: Irrational Case\label{DLCI}}

Let us now examine the general form of Morita equivalent loop
correlators that arises when the noncommutativity parameter
$\Theta$ is an irrational number. In this case, the target theory is
necessarily another noncommutative gauge theory. This dual gauge
theory is once again defined on a torus whose size $\tilde r$ depends
on the noncommutativity parameter as prescribed in
(\ref{Moritagen}). This means that as we go from our original
noncommutative gauge theory to its Morita dual, the size of the torus
may change drastically. More precisely, we recall that to every Morita
equivalence parameterized by $SL(2,\zed)$ integers $a,b,c,d$, there
exists a critical radius $r_{\rm c}^{a,b}$ given by (\ref{rcritdef})
which is associated to each path such that if the radius of
the target torus $\tilde r$ is smaller than $r_{\rm c}^{a,b}$, then
the path can self-intersect in the dual gauge theory. Whether or not
self-intersections actually occur depends on the shape of the path
itself, as well as on its width, length and orientation. The key
point is that, if the noncommutativity parameter is irrational-valued,
then the critical radius $r_{\rm c}^{a,b}$ can be made
vanishingly small. This is a consequence of the well-known number
theoretic property~\cite{Hardy1} that, given $\Theta \in \mathbb{R
  \,\backslash\,Q}$, the subset $\mathbb{Z} + \mathbb{Z}\,\Theta = \{
a + b\,\Theta\,|\, a,b \in\mathbb{Z} \}$ is dense on the real line
$\mathbb{R}$. In particular, given any $\varepsilon > 0$, we can
always find $a,b \in\mathbb{Z}$ such that $|a +
b\,\Theta|<\varepsilon$ and hence $r_{\rm c}^{a,b}<\varepsilon\, r
$. In other words, since the area of a closed Wilson line does not
change under a Morita transformation, any Wilson loop in irrational
noncommutative Yang--Mills theory is dual to a Wilson loop with
arbitrarily many self-intersections and windings around the
torus. This gives a combinatorial picture of irrational Wilson loops
as densely wound and interesecting loops on arbitrarily small tori.

On more heuristic grounds, we can rephrase our argument as
follows. Let us take $\Theta\in\real\,\backslash\,\rat$ and
approximate it by a sequence of rational numbers as
\begin{equation}
\Theta = -\lim_{n \to \infty}\,\frac{c_n}{N_n} \ ,
\end{equation}
where both sequences of integers $c_n$ and $N_n$ tend to infinity such
that their ratio is held fixed in the limit. For every fixed $n$, we
can choose a Morita transformation such that $r_{\rm c}^{a_n,b_n} = r/N_n$.
In the limit $n \to \infty$, one has $r_{\rm c}^{a_n,b_n}\to 0$. With
$\ell^\mu(\mathcal{C})$ the characteristic lengths associated with the
path $\mathcal{C}$ which we introduced in (\ref{charlengths}), it
follows that it is always possible to find a bound of the form
(\ref{bound}). In other words, whenever $\Theta$ is an irrational
number, there is a target torus on which the given path
self-intersects and winds an (uncountably) infinite number of
times. Recalling the analysis of the previous section, we see that the
apparent violation of invariance under area-preserving diffeomorphisms
is in fact due to the self-intersecting nature of dual Wilson
loops. Differently shaped loops can have drastically different
self-intersection and winding images under the same Morita
transformation.

This self-intersecting property presents a serious technical
obstruction to obtaining exact nonperturbative expressions for
correlation functions of irrational noncommutative Wilson loops. In
particular, the loop functional is not a smooth function of $\theta$,
and the geometrical path parameters display a drastic change under
Morita equivalence. It is thus not clear what a Morita
duality-invariant expression for closed Wilson line correlators should
look like. However, there is a natural and obvious regime in which
exact results can be obtained. If one considers a certain double
scaling limit in which the area enclosed by the Wilson loop vanishes
faster than the area of the target torus, then the dual Wilson
loop will be of the same (non-intersecting) type. This particular
limit is the topic of the next section.

\section{Loop Correlators in the Double Scaling Limit \label{Wilsoninf}}

In this section we will compute a particular class of noncommutative
loop correlators that can be consistently obtained through the use of
Morita equivalence. Consider a loop winding $n$ times around
itself and encircling an area $\rho_1$. As in Section~\ref{Simple},
the area outside the loop is denoted $\rho_2$ so that the total area
of the torus is $(2\pi\,r'\,)^2=\rho_1+\rho_2$. Having in mind the
picture of noncommutative Wilson loops drawn out in the previous
section, we will take the limit
$n\to\infty$ with the product $n^2\,\rho_1=\lambda$ held fixed. Because the
loop area can be taken arbitrarily small in the $\tilde r\to0$ limit
required to induce gauge theory on the noncommutative
plane~\cite{GSV1}, the final result should be consistent with the
known expression obtained by resumming the small loop area
perturbation series on $\real^2$~\cite{BNT}.

Our starting point is the general expression (\ref{path1U(N)}) for the
Wilson loop correlator in the $k^{\rm th}$ topological sector of the
dual $SU(N)$ gauge theory on $\torus^2$. For the representation $R$ we
take $R=N^{\otimes n}$ which, since $\chi^{~}_{N^{\otimes
    n}}(U)=\chi^{~}_N(U^n)$, describes a Wilson loop in the
fundamental representation with $n$ windings. We will compute the
corresponding normalized correlation function
\beq
\mathcal{W}_n^k(\rho_1)=\frac{W_{\mathcal{C}_1;N^{\otimes
      n}}^k(\rho_1)}{N\,Z_k}
\label{Wgennorm}\eeq
where
\beq
Z_k=\sum_R\e^{-\frac{g^2\,(\rho_1+\rho_2)}2\,C_2(R)}~\chi^{~}_R
\left(\e^{2\pi\ii k/N}\right)
\label{partfnk}\eeq
is the partition function of Yang--Mills theory on the torus in the
$k^{\rm th}$ 't~Hooft sector. As in (\ref{path1U(N)final}), the required
Clebsch--Gordan coefficients can be computed from the explicit
expression
\bea
\mathrm{N}^{R_2}_{~~R_1N^{\otimes n}}&=&
\prod_{a=1}^N\,\int_0^1 \mathrm{d} \lambda_a~
\delta\left(\,\mbox{$\sum\limits_{a=1}^N$}\,\lambda_a\right)~\sum_{c=1}^N
\e^{2 \pi \ii n\,\lambda_c}\nonumber\\ && \qquad \times\,
\det_{1 \le a,b \le N} \left[\e^{2
\pi \ii n^{R_1}_a\,\lambda_b} \right]\,\det_{1 \le a,b \le N}
\left[\e^{2 \pi \ii n^{R_2}_a\,\lambda_b} \right] \ .
\label{CGcoeffn}\eea
It is convenient to introduce integers $l^{R_i}_N$, $i=1,2$
through the identities
\begin{equation}
1=\frac1{\sqrt\pi}\,\int_0^1\dd\alpha_i~
\sum_{l^{R_i}_N=-\infty}^{\infty}\e^{-(2\pi)^2\,\bigl(\alpha_i
-\frac1N\,\sum\limits_{a=1}^{N-1} n^{R_i}_a -l^{R_i}_N\bigr)^2} \ ,
\end{equation}
and to change summation variables from Young tableau boxes to integers
$l^{R_i}_a$, $a=1,\dots,N-1$ and $l^{R_i}$ defined by~\cite{BGV}
\beq
l^{R_i}_a=n^{R_i}_a  + l^{R_i}_N -a+N \ ,
\quad  l^{R_i}=\sum_{a=1}^N\,l^{R_i}_a \ .
\eeq
In terms of these new integers, the quadratic Casimir invariant and
dimension of the representation $R_i$ are given by
\beq
\label{casimiri} C_2\left(R_i\right)=C_2\left(\mbf l^{R_i}\right)
= \sum_{a=1}^{N} \left(l^{R_i}_{a}-\frac{l^{R_i}}{N}
\right)^2- \frac{N}{12}\,\left(N^2-1\right) \ , \quad
\dim R_i=\Delta\left(\mbf l^{R_i}\right) \ .
\eeq

By exploiting the complete symmetry of the correlator in the summation
integers we thereby arrive at the expression
\begin{eqnarray}
\mathcal{W}^{k}_n(\rho_1)&=&\frac{1}{N\,Z_k}\,\frac1{(2\pi)^N\,\pi\,
(N!)^2}\,\prod_{a=1}^N\,\int_0^1 \mathrm{d} \lambda_a~
\delta\left(\,\mbox{$\sum\limits_{a=1}^N$}\,\lambda_a\right)~\sum_{c=1}^N
\e^{2 \pi \ii n\,\lambda_c}\nonumber\\ &&\times\,
\sum_{\mbf l^{R_1},\mbf l^{R_2}}\,\frac{\Delta\left(\mbf
    l^{R_1}\right)}{\Delta\left(\mbf l^{R_2}\right)}~
\e^{-\frac{g^2\,\rho_1}{2}\,C_2(\mbf
  l^{R_1})-\frac{g^2\,\rho_2}{2}\,C_2(\mbf l^{R_2})}~
\e^{2\pi\ii k\,l^{R_1}/N}~\prod_{a=1}^N
\e^{2\pi\ii\bigl(l_a^{R_2}-l_a^{R_1}\bigr)\,\lambda_a}\nonumber\\
&&\times\,\int_0^1\dd\alpha_1~\e^{-(2\pi)^2\,
\bigl(\alpha_1 - \frac{l^{R_1}}N  \bigr)^2}~\int_0^1
\dd\alpha_2~\e^{-(2\pi)^2\,\bigl(
\alpha_2 - \frac{l^{R_2}}N \bigr)^2 } \ .
\end{eqnarray}
Expressing the delta-function as a Fourier series and integrating over
$\alpha_1$, we convert this expression into the form
\begin{eqnarray}
\mathcal{W}^{k}_n(\rho_1)&=&\frac{1}{N\,Z_k}\,\frac1{(2\pi)^N~\sqrt\pi~
(N!)^2}\,\prod_{a=1}^N\,\int_0^1 \mathrm{d} \lambda_a~\sum_{c=1}^N
\e^{2 \pi \ii n\,\lambda_c}\nonumber\\ &&\times\,
\sum_{\mbf l^{R_1},\mbf l^{R_2}}\,\frac{\Delta\left(\mbf
    l^{R_1}\right)}{\Delta\left(\mbf l^{R_2}\right)}~
\e^{-\frac{g^2\,\rho_1}{2}\,C_2(\mbf
  l^{R_1})-\frac{g^2\,\rho_2}{2}\,C_2(\mbf l^{R_2})}~
\e^{2\pi\ii k\,l^{R_1}/N}~\prod_{a=1}^N
\e^{2\pi\ii\bigl(l_a^{R_2}-l_a^{R_1}\bigr)\,\lambda_a}\nonumber\\
&&\times\,\int_0^1\dd\alpha~\e^{-(2\pi)^2\,
\bigl(\alpha - \frac{l^{R_2}}N  \bigr)^2} \ .
\end{eqnarray}
We use the complete symmetry again to now fix the summation index $c=1$,
which produces an additional factor of $N$. The $\lambda_a$ integrals can now
be
performed explicitly giving the constraints $l^{R_1}_a=l^{R_2}_a$,
$a\neq 1$ and $l^{R_1}_1=l^{R_2}_1+n$. This leads to our final
explicit result
\begin{eqnarray}
\mathcal{W}^{k}_n(\rho_1)&=&\frac{1}{\sqrt\pi~N!~Z_k}~
\e^{-\frac{g^2\,n^2\,\rho_1}{2}\,\left(1-\frac{1}{N}\right)}~\sum_{\mbf
n}\,\frac{\Delta(n_1+n,n_2,\dots,n_N)}
{\Delta(\mbf n)}~\e^{-\frac{g^2\,(\rho_1+\rho_2)}{2}\,C_2(\mbf n)} \nonumber\\
&&
\times~\e^{-\frac{g^2\,
\rho_1}{2}\,\bigl(2n\,n_1-\frac{2n}{N}\,\sum\limits_{a=1}^Nn_a\bigr)}~
\e^{\frac{2\pi\ii k}{N}\,\sum\limits_{a=1}^Nn_a}~
 \int_0^1\dd\alpha~\e^{ -(2\pi)^2\,\bigl(\alpha-
   \frac1N\,\sum\limits_{a=1}^Nn_a\bigr)^2} \ .
\label{5.1}\end{eqnarray}
We can check the normalization here by observing that these same steps
can be used to write the partition function (\ref{partfnk}) as
\begin{equation}
Z_k=\frac{1}{\sqrt{\pi}~ N!}\,\sum_{\mbf
  n}\e^{-\frac{g^2\,(\rho_1+\rho_2)}
{2}\,C_2(\mbf n)}~\e^{\frac{2\pi\ii k}{N}\,\sum\limits_{a=1}^Nn_a}~
\int_0^1\dd\alpha~\e^{ -(2\pi)^2\,\bigl(\alpha -
  \frac1N\,\sum\limits_{a=1}^Nn_a\bigr)^2} \ .
\end{equation}
Thus our conventions imply the normalization condition
$\mathcal{W}^{k}_0(\rho_1)=1$.

Let us now take the $n\to\infty$ limit. For this, we insert the
explicit expression for the Vandermonde determinant (\ref{dimRVan}) to
recast (\ref{5.1}) as
\begin{eqnarray}
\mathcal{W}^{k}_n(\rho_1)&=&\frac{1}{\sqrt\pi~N!~Z_k}~\sum_{\mbf
n}\e^{-\frac{g^2\,(\rho_1+\rho_2)}{2}\,C_2(\mbf n)}~
\e^{\frac{2\pi\ii k}{N}\,\sum\limits_{a=1}^Nn_a}~
\int_0^1\dd\alpha~\e^{ -(2\pi)^2\,\bigl(\alpha-
   \frac1N\,\sum\limits_{a=1}^Nn_a\bigr)^2}
\nonumber\\ &&\times~\e^{-\frac{g^2\,
\rho_1}{2}\,\bigl(n^2\,(1-\frac{1}{N})+2n\,n_1-\frac{2n}{N}\,
\sum\limits_{a=1}^Nn_a\bigr)}~
\sum_{m=0}^{N-1}\,\frac{(N-1)!}{(N-1-m)!}\,\prod_{j=2}^{m+1}\,\frac{n^m}{n_1-n_j}
\ .
\end{eqnarray}
We thus obtain a Laurent series in $\frac1n$ from expanding the
exponential term to get
\begin{eqnarray}
\mathcal{W}^{k}_n(\rho_1)&=&\frac{1}{\sqrt\pi~N!~Z_k}~\sum_{\mbf
n}\e^{-\frac{g^2\,(\rho_1+\rho_2)}{2}\,C_2(\mbf n)}~
\e^{\frac{2\pi\ii k}{N}\,\sum\limits_{a=1}^Nn_a}~
\int_0^1\dd\alpha~\e^{ -(2\pi)^2\,\bigl(\alpha-
   \frac1N\,\sum\limits_{a=1}^Nn_a\bigr)^2}\nonumber\\ && \qquad
\times\,\sum_{l=0}^{\infty}~\sum_{m=0}^{N-1}~\sum_{p=0}^{l}\,
\frac{(N-1)!}{(N-1-m)!}~n^{m-l}~\frac{\left(-g^2\,\lambda
\right)^l}{p!\,(l-p)!}\nonumber\\ && \qquad \times\,\left(-\frac{1}{N}\,
\sum_{a=1}^Nn_a^2\right)^{l-p}~
\prod_{j=2}^{m+1}\,\frac{n_1^p}{n_1-n_j}~\e^{-\frac{g^2\,\lambda}{2}\,
\left(1-\frac{1}{N}\right)} \label{5.2}
\end{eqnarray}
where $\lambda=n^2\,\rho_1$. The very same structure appears in the
computation of $n$-winding Wilson loops on the sphere~\cite{BGV} and
it is clear that this result generalizes to arbitrary genus Riemann
surfaces. Let us thus proceed as in~\cite{BGV}.

First of all, we observe that (\ref{5.2}) is actually an expansion in
$\frac1{n^2}$. This point can be understood by changing $\mbf
n\to-\mbf n$, which produces an overall factor $(-1)^{m-l}$ weighting
the sum over $\mbf n$. This implies that $m-l$
must be an even integer in order to contribute a non-vanishing
result. It is useful to now rewrite the sum over $\mbf n$ for fixed
$l,m,p$ as
\begin{equation}
\frac1{(m+1)!}\,\sum_{\mbf n\in\zed^N} \e^{-
\frac{g^2\,(\rho_1+\rho_2)}2\,\sum\limits_{a=1}^N n_a^2}\,
\left(-\frac{1}{N}\,\sum_{a=1}^N
n_a^2\right)^{l-p}\,\sum _{\pi\in S_{m+1}}~\prod_{j=2}^{m+1}\,
\frac{n_{\pi(1)}^p}{n_{\pi(1)}-n_{\pi(j)}} \ . \label{5.3}
\end{equation}
Let us evaluate the zeroth-order contribution to (\ref{5.2}). For
$l=p=m$ one can write the sum over permutations as
\begin{equation}
\sum_{\pi\in
S_{m+1}}\,\prod_{j=2}^{m+1}\,\frac{n_{\pi(1)}^m}{n_{\pi(1)}-n_{\pi(j)}} =
\frac1{\Delta(n_1,\dots,n_{m+1})}\, \sum_{\pi\in S_{m+1}}
f_\pi(n_1,\dots,n_{m+1}) \ ,
\label{permsum}\end{equation}
where the Vandermonde determinant arises from the common
denominator and the quantity $\sum_{\pi\in S_{m+1}}\,
f_\pi(n_1,\dots,n_{m+1})$ is a polynomial of degree
$\frac12\,m\,(m+1)$ in $m+1$ variables. Since the Vandermonde
determinant $\Delta(n_1,\dots,n_{m+1})$ is completely antisymmetric in
its arguments, the non-vanishing contribution to (\ref{5.3}) comes
from the completely antisymmetric part of $\sum_{\pi\in S_{m+1}}\,
f_\pi(n_1,\dots,n_{m+1})$ implying that
\begin{equation}
\sum_{\pi\in S_{m+1}}
f_\pi(n_1,\dots,n_{m+1})=C~\Delta(n_1,\dots,n_{m+1}) \ .
\end{equation}
The proportionality constant is easily found by inspection to be
$C=1$. It is not difficult to prove that the potentially divergent
contributions in the limit $n\to\infty$, coming from the terms with
$l<m$ in (\ref{5.2}), vanish. For this, we again use
(\ref{5.3}) to notice that we can still factorize a
Vandermonde determinant in the denominator as in (\ref{permsum}), but
now $\sum_{\pi\in S_{m+1}}\,f_\pi(n_1,\dots,n_{m+1})$ is a polynomial
of degree less than $\frac12\,m\,(m+1)$ in $m+1$ variables because
$p<m$. Its completely antisymmetric part thus vanishes.

In this way we arrive finally at
\begin{equation}
\mathcal{W}_\infty\left(g^2\,\lambda\right)
=\lim_{n\to\infty}\,\mathcal{W}^{k}_n\left(
\mbox{$\frac\lambda{n^2}$}\right)=\e^{-\frac{g^2\,
\lambda}{2}\,\left(1-\frac{1}{N}\right)}~
\sum_{m=0}^{N-1}\,\frac{(N-1)!}{(N-1-m)!}\,\frac{\left(-g^2\,\lambda
\right)^m}{m!\,(m+1)!} \ .
\end{equation}
Corrections to this formula are of order $\frac1{n^2}$. Note that the
partition function cancels in this limit. The Wilson loop average can
be expressed in terms of a generalized Laguerre polynomial as
\begin{equation}
\mathcal{W}_\infty\left(g^2\,\lambda\right)=
\mbox{$\frac{1}{N}$}~\e^{-\frac{g^2\,
\lambda}{2}\,\left(1-\frac{1}{N}\right)}~
L^1_{N-1}\left(g^2\,\lambda\right) \ .
\end{equation}
Rather remarkably, this result coincides with the analogous result for
Yang--Mills theory on the sphere. In fact, it is completely
independent of the genus of the original Riemann surface. Differences
would appear only at sub-leading order in $\frac1{n^2}$.

At this point we can take the large $N$ limit to reach gauge theory on
the noncommutative plane. The noncommutative Yang--Mills coupling
constant in this case is defined through $g^2=\Theta\,\hat g^2=\hat
g^2/N$, and the Wilson loop correlator can be expressed in terms of a
Bessel function as
\begin{equation}
\hat{\mathcal W}_\infty\left(\hat g^2\,\lambda\right)=
\lim_{N\to\infty}\,\mathcal{W}_\infty\left(\mbox{$\frac{\hat
      g^2\,\lambda}N$}\right)=\frac{J_1\left(2\,\sqrt{\hat{g}^2\,\lambda}
\,\right)}{\sqrt{\hat{g}^2\,\lambda}} \ .
\end{equation}
This expression coincides exactly with the result obtained, at this
order, by resumming the perturbation series~\cite{BNT}. The
coincidence of the correlator of the noncommutative Wilson loop in the
present limit with that of the commutative Wilson loop obtained by
resumming planar diagrams confirms the general expectation~\cite{BHN1} that
noncommutativity modifies only large area Wilson loops. For small
loops the usual commutative behaviour at large $N$ is recovered, while
large area loops become complex-valued~\cite{BHN1}. The double scaling limit we
have considered in this section effectively singles out small area
loops. The fact that noncommutativity only modifies the long
wavelength behaviour of Wilson loops is indicative of some
nonperturbative form of UV/IR mixing. This mixing only affects the
closed Wilson line observables of the noncommutative gauge theory and
is another manifestation of the loss of invariance under
area-preserving diffeomorphisms. Note that the present double scaling
limit ``zooms in'' on only a very small portion of the torus, so that
the final correlator in the limit is completely independent of any
global properties of the two-dimensional spacetime. This small area
limit is equivalent to the limit $\theta=\infty$, as one might have
naively expected, and therefore eliminates all higher order traces of
the $\frac1\theta$-expansion.

\section{Summary and Discussion\label{Summary}}

In this paper we have explored new aspects of the shape dependence of
Wilson loop correlators on a two-dimensional noncommutative
torus. Because of the non-trivial topology and the compactness of the
spacetime, correlation functions associated to loops $\mathcal{C}$ of
the same area apparently depend not only on their shape but also on
their characteristic
lengths $\ell^\mu(\mathcal{C})$ defined in (\ref{charlengths})
(i.e. their heights and widths) and on their orientation in the torus. We
illustrated this dependence through several explicit calculations
using Morita equivalence along with a combinatorial approach. From our
perspective the observed breaking of invariance of Wilson loop
correlators under area-preserving diffeomorphisms of the
two-dimensional spacetime may be attributed to the wrapping and
self-intersecting nature of Morita dual Wilson loops. Only those
contours whose images under Morita equivalence lead to isomorphic
non-planar graphs will give rise to identical correlation
functions. In irrational noncommutative gauge theory, there always
exist dual loops which wind infinitely many times around the
torus. Motivated by this picture, we have also explicitly computed an
infinitely wound Wilson loop correlator. Since the limit of infinite
winding considered corresponds to small loop area, our results agree
with those obtained by resumming planar diagrams in {\it commutative}
gauge theory, the planarity arising essentially as a combined large
$N$ and large $\theta$ effect. This limit also eliminates the
non-perturbative topological degrees of freedom which are expected to
restore the usual large $N$ Gross--Witten area law
behaviour~\cite{GW1} for small noncommutative Wilson loops.

It is interesting to examine in the present context the perturbative
anomaly that comes from the contribution of the non-planar diagram of
order $\hat g^4$ to the average of the noncommutative Wilson loop on
$\real^2$~\cite{ADM1}. The leading term in the $\frac1\theta$ expansion of the
correlator is proportional to $\hat g^4\,\rho_1^2$ in our notation, and
thus it survives the limit $\theta\to\infty$ due to the singular infrared
behaviour of the gauge propagator in two dimensions. This term appears
to be in conflict with both general arguments of noncommutative
perturbation theory~\cite{MVRS1} and with the representation of noncommutative
gauge theories via large $N$ twisted reduced models~\cite{AMNS1}. However, the
anomalous term vanishes in the double scaling limit considered in
Section~\ref{Wilsoninf} and so does not show up in our calculations
which capture the entire small loop area perturbation series.

There may be yet another way to eliminate this anomalous behaviour.
We offer the following argument only as a somewhat speculative
conjecture at this stage. The perturbative calculations~\cite{ADM1} which
unveil this anomalous term are performed in the axial gauge where the
self-interactions of the gauge field disappear. As is well-known, the
axial gauge is forbidden on the torus (or on any spacetime of non-trivial
topology) due to the existence of topologically non-trivial field
configurations (transforming under large gauge transformations) which
yield non-trivial Polyakov loops along the axial direction. In
commutative gauge theory on $\real^2$ only topologically trivial gauge
fields exist (transforming under gauge transformations connected to
the identity) and there is no problem with the axial gauge
choice. However, this is {\it not} the case for gauge theory on the
noncommutative plane. In contrast to the commutative situation the
gauge theory now contains topologically non-trivial backgrounds called {\it
  fluxons}~\cite{Poly1,GrossNek1} owing to the fact that
$\real_\theta^2$ has, like the torus, a non-trivial K-theory
group. The fluxons can be regarded~\cite{GSS1} as the surviving
degrees of freedom left over from the usual instanton configurations
in the limit where one decompactifies the noncommutative torus onto
the noncommutative plane. An $L$-fluxon solution is labelled by a
set of moduli $\mbf\lambda_1,\dots,\mbf\lambda_L\in\real^2$, which
specify the locations of the vortices on the plane, and by a
collection of magnetic charges $m_1,\dots,m_L\in\zed$. One can compute the
semi-classical average of an {\it open} Wilson line operator along a
straight infinite contour pointing in a direction $\hat{\mbf e}$ of
$\real^2$ in the fluxon background with the result~\cite{GrossNek1}
$W_{\rm open}(\hat{\mbf e})=\sum_{a=1}^L\e^{\ii\hat{\mbf
    e}\cdot\mbf\lambda_a}$ (independently of the vortex
charges). Generically, this expectation value cannot
be trivialized by any noncommutative gauge transformation and the
correlator thus presents an obstruction to choosing the axial
gauge. Axial gauge choices are also forbidden in the lattice
regularization of noncommutative Yang--Mills theory due to UV/IR
mixing~\cite{PanSz3}. This fact suggests that the observed anomalous
behaviour of noncommutative Wilson loops could be due to the choice of a wrong
vacuum, and that the correct perturbative calculation should instead
expand about the background of a fluxon. It would be interesting
to investigate this point further.

\acknowledgments

We thank A.~Bassetto, Y.~Makeenko and F.~Vian for helpful
discussions. The work of M.C. and R.J.S. was supported in part by
PPARC Grant PPA/G/S/2002/00478 and by the EU-RTN Network Grant
MRTN-CT-2004-005104. The work of R.J.S. was
supported in part by a PPARC Advanced Fellowship.

\end{document}